\documentclass[12pt]{iopart}
\usepackage{amsmath, amsfonts, amssymb, amsthm, amstext, bm, bbm, multirow, url, comment}
\usepackage{lmodern}
\usepackage[T1]{fontenc}
\usepackage{braket}
\usepackage{graphicx, subcaption, ragged2e}
\DeclareCaptionJustification{justified}{\justifying}

\usepackage[usenames, dvipsnames]{xcolor}
\usepackage[colorlinks=true, linkcolor=blue,citecolor=red, urlcolor=blue, hypertexnames=true]{hyperref}
\usepackage[textwidth=160mm, textheight = 210mm, hmarginratio={1:1}, vmarginratio={1:1}, 
			 heightrounded]{geometry} 
\usepackage[version=4]{mhchem}

\renewcommand{\theequation}{\thesection.\arabic{equation}}

\usepackage{color}
\definecolor{darkblue}{rgb}{0,0,0.6}
\definecolor{darkred}{rgb}{0.6,0,0}
\definecolor{darkgrey}{rgb}{0.6,0.6,0.6}

\newcommand{\stkout}[1]{\ifmmode\text{\sout{\ensuremath{#1}}}\else\sout{#1}\fi}

\begin{document}

\title{Exact Fluctuating Hydrodynamics\\ of the Scaled Light-Heavy Model}

\author{Shilpa Prakash}
\address{Tata Institute of Fundamental Research, Hyderabad, 500046, India}
\author{Mustansir Barma}
\address{Tata Institute of Fundamental Research, Hyderabad, 500046, India}
\author{Kabir Ramola}
\address{Tata Institute of Fundamental Research, Hyderabad, 500046, India}

\begin{abstract}
We study the exact fluctuating hydrodynamics of the scaled Light-Heavy model (sLH), in which two species of particles (light and heavy) interact with a fluctuating surface. This model is similar in definition to the unscaled Light-Heavy model (uLH), except it uses rates scaled with the system size. The consequence, it turns out, is a phase diagram that differs from that of the unscaled model. We derive the fluctuating hydrodynamics for this model using an action formalism involving the construction of path integrals for the probability of different states that give the complete macroscopic picture starting from the microscopic one. 
This is then used to obtain the two-point steady-state (static) correlation functions between fluctuations in the two density fields in the homogeneous phase. We show that these theoretical results match well with microscopic simulations away from the critical line.
We derive an exponentially decaying form for the two-point steady-state correlation function with a correlation length that diverges as the critical line is approached.
Finally, we also compute the dynamic correlations in the homogeneous phase and use them to determine the relaxation dynamics as well as the dynamic exponents of the system. 


	\end{abstract}	
\maketitle

\tableofcontents

\section{Introduction}
\label{sec:intro}
Interacting non-equilibrium systems have been known to present a host of interesting phenomena, such as coarsening, phase separation and dynamical arrest, all of which find relevance in diverse fields such as physics and biology~\cite{Fang_2019, Gnesotto2018}. Archetypical of this are driven granular materials~\cite{Jaeger_1996}, turbulent mixtures~\cite{Berti_2005}, constrained systems at low temperatures~\cite{Testard_2014}, as well as soft matter and biological systems. Although the most general hardcore particle systems known to display glassy dynamics leading up to phase separation are hard to characterize theoretically, one can still gauge much information from simple models of confined particles on the lattice. In the thermodynamic limit, hydrodynamic laws can describe lattice models — this works by transitioning from the microscopic to the macroscopic level by suitable rescaling of time and space. The rescaling results from many particles and microscopic units of length getting squeezed into a single macroscopic unit. Similarly, many elementary events across multiple microscopic intervals are added, giving us an eventually blurred macroscopic picture.

In recent years, there has been much progress in developing hydrodynamic equations for systems that exhibit stochastic dynamics~\cite{Derrida_2007}. However, systems with interactions do pose some challenges. Hardcore interactions, for example, are known to produce dynamic correlations~\cite{Lebowitz_1988}.~A workaround for this complexity is fortunately possible, since even with hardcore interactions, steady states are simple to characterize. Consequently, one can easily arrive at the hydrodynamics of these systems by scaling the steady-state particle current. These models in which the rates are rescaled in a particular way are amenable to an exact hydrodynamic description. The ABC model, for example, has been studied in both the unscaled limits (uABC)~\cite{Evans_1998} and the scaled (sABC)~\cite{Clincy_2003} limits and solved exactly in the latter, where a phase transition was observed. Both limits of the Light-Heavy model, which we will be dealing with, have also been analyzed, with the studies primarily focused on the unscaled limit (uLH)~\cite{Lahiri_1997, Lahiri_2000,dasbasu_2001, ramaswamy_2002, Mahapatra2020Light, Khamrai_2024,Chakraborty_2016,Chakraborty2_2017,Chakraborty_2019}. In this paper, we will be focusing on the scaled regime, which we shall refer to as the scaled Light-Heavy (sLH) model. 

The primary difference in the behavior of the scaled and unscaled models is due to the microscopic dominance of diffusion in the former.~As diffusion dominates on a local scale in such systems, this allows one to neglect correlations on lengthscales beyond a single coarse-graining box. However, the bias becomes relevant at large length scales, and therefore is a crucial parameter in the description of such systems at the hydrodynamic scale~\cite{Derrida_2007, martin2021}. Consequently, working with scaled models enables an exact formulation of hydrodynamics for such systems. The deterministic nature of hydrodynamics is lost due to fluctuations that are present in finite-sized systems.~To take this into account, one must explicitly include the fluctuation terms in equations describing such systems; these modified equations with added noise terms then describe the fluctuating hydrodynamics of the system. In the scaled models under consideration, the correlations between these fluctuations are purely diffusive in nature. Techniques, such as the Doi-Peliti method, allow one to derive the exact form of fluctuating hydrodynamics starting from the microscopic description in several of these models. Recently, Agranov {\it et al.} analyzed the steady-state properties of a lattice gas model of active particles using the fluctuating hydrodynamic framework~\cite{agranov2021exact}. This Active Lattice Gas (ALG) model, defined on a 1D ring lattice, has particles of the Ising variety moving under the effect of three processes: diffusion, bias, and tumbling, or stochastic switching between different directions of motion. Since the rates of these processes are suitably scaled, analytic expressions for the static and dynamic correlations can be derived for this model by considering fluctuations in density up to linear order about the homogeneous steady-state~\cite{agranov2021exact}.

The scaled Light-Heavy model (sLH) resembles the active lattice gas model in the hydrodynamic limit (except for a non-linear noise term and the nature of noise correlations) and yet remains distinct when examined at the microscopic level. The Light-Heavy model was first introduced
by Lahiri and Ramaswamy in the unscaled regime (uLH) in~\cite{Lahiri_1997} and was explored in detail in \cite{ Lahiri_2000, dasbasu_2001, ramaswamy_2002, Mahapatra2020Light, Khamrai_2024}.
It consists of hard-core particles that interact with a fluctuating surface, following coupled dynamics governed by the rules illustrated in Figure~\ref{pic2}a. Although initially introduced to study instabilities in sedimentation~\cite{Ramaswamy_2001}, the model finds relevance in more general systems that exhibit coupled dynamics.
For example, proteins and lipids on cell membranes cluster due to fluctuations in the cell membrane~\cite{goswami2008nanoclusters, garcia2014nanoclustering}. Passive models for such a system display dynamic cluster formation and disintegration~\cite{das_2016}. Recent experiments~\cite{varma2006tcell, yu_2011} show proteins and lipids influence the membrane, affecting clustering in ways passive models miss, highlighting the need for coupled systems like the LH model.

Particles and surface units, both described by Ising variables, exhibit coupled dynamics leading to diffusive motion of the constituents with a fluctuating bias, and this two-way coupling determines the kind of organization that will emerge. This results in a rich phase diagram with multiple phases. When considering the hydrodynamic limit, the scaling of the bias parameters with system size results in nontrivial diffusive scaling, shifting the model to a new class with a phase diagram that is distinct from the unscaled version. 
The detailed characterization of the dynamic and static properties of the different ordered and disordered phases in the uLH model was carried out in~\cite{ Chakraborty2_2017, Chakraborty_2017}, with Ref.~\cite{Chakraborty_2017} also exploring a phase transition observed in the sLH model. In this work, we focus on the disordered phase of the scaled model. Specifically, we are interested in the validity of a hydrodynamic description as well as the correlation between density fluctuations of the different fields in the homogeneous state.
In this region of the phase diagram, we find that the model is characterized by a length scale, which grows as the critical point is approached, indicating emerging collective behavior. We predict that the correlation length specifically displays a square-root divergence with respect to the distance to the critical point.
We also perform Monte Carlo simulations on this model, and verify our theoretical predictions.


The outline of this paper is as follows: in Section~\ref{sec:model}, we describe the scaled and unscaled models and recall their hydrodynamics, which was derived in Ref.~\cite{Chakraborty_2017}. In Section~\ref{sec:Action-LH}, we obtain the fluctuating hydrodynamics for the scaled model, which we derive using a path integral formulation. Section~\ref{sec:derivation-two-point} describes our results, the analytically computed two-point correlation for different density fields. 
We validate our theoretical results using numerical simulations of the lattice model.
Section~\ref{sec: phase transition} discusses a phase transition in the scaled model. Finally, in Section~\ref{sec:discussion}, we conclude and provide directions for further research.

 \section{The model and its noiseless hydrodynamics}
\label{sec:model}
\begin{figure}[t!]
\centering
    \includegraphics[width=1\linewidth]{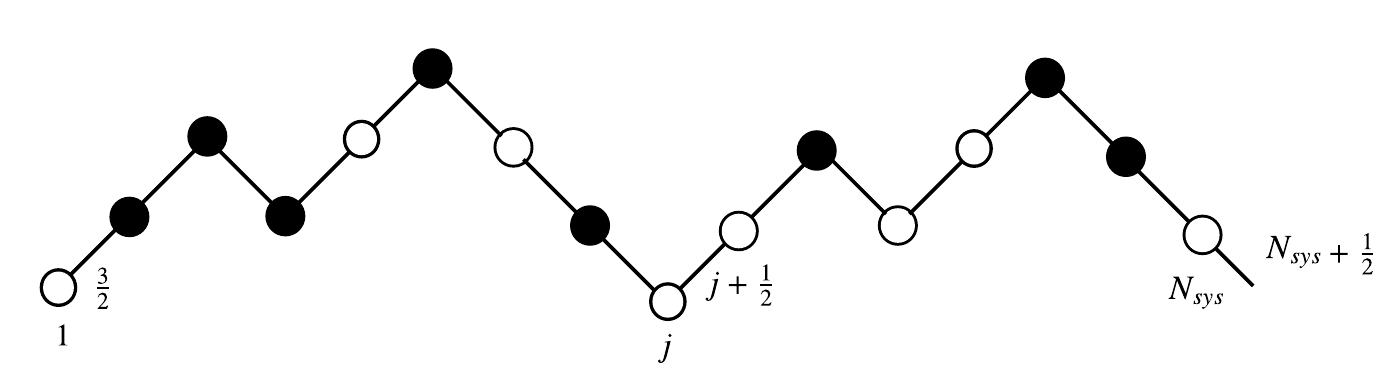}
  \caption{A schematic of a configuration of the Light-Heavy model. Light particles ($\circ$) and heavy particles ($\bullet$) are placed at integer sites $j$ on a lattice, while up tilts ($/$) and down tilts ($\setminus$) are located at half integer sites $j + \frac{1}{2}$. }
  \label{schem1}
\end{figure}

The Light-Heavy (LH) model is a lattice system of particles and tilts with coupled dynamics. 
In this model, the two species occupy separate sublattices, each
allowed to take up one of two states, as shown in Figure~\ref{schem1}. Here, we consider a one-dimensional system. Particles can be either heavy or light, while tilts
can have up or down slopes. Using variables $\sigma$ and $\tau$, label the two species with locations
indexed by $j$; for example, we have chosen the following convention here.
\vspace{3mm}
\newline
$\sigma_{j} = 
        \begin{array}{cc}
         \Biggl\{ &
          \begin{matrix}
           -1 & \text{Light}  &\circ\\
           +1  & \text{Heavy}  &\bullet    
          \end{matrix}
        \end{array}\hspace{4cm}
\tau_{j+\frac{1}{2}} = 
        \begin{array}{cc}
         \Biggl\{ &
          \begin{matrix}
           -1  &\text{Up}  &/\\
           +1  &\text{Down} &\setminus    
          \end{matrix}
        \end{array}$
\vspace{3mm}
\newline
The system is assumed to have periodic boundary conditions. Consequently, the average slope is $0$. Additionally, each sublattice has an even number of sites.

\begin{figure}[t!]
\centering
\includegraphics[width=1.0\linewidth] {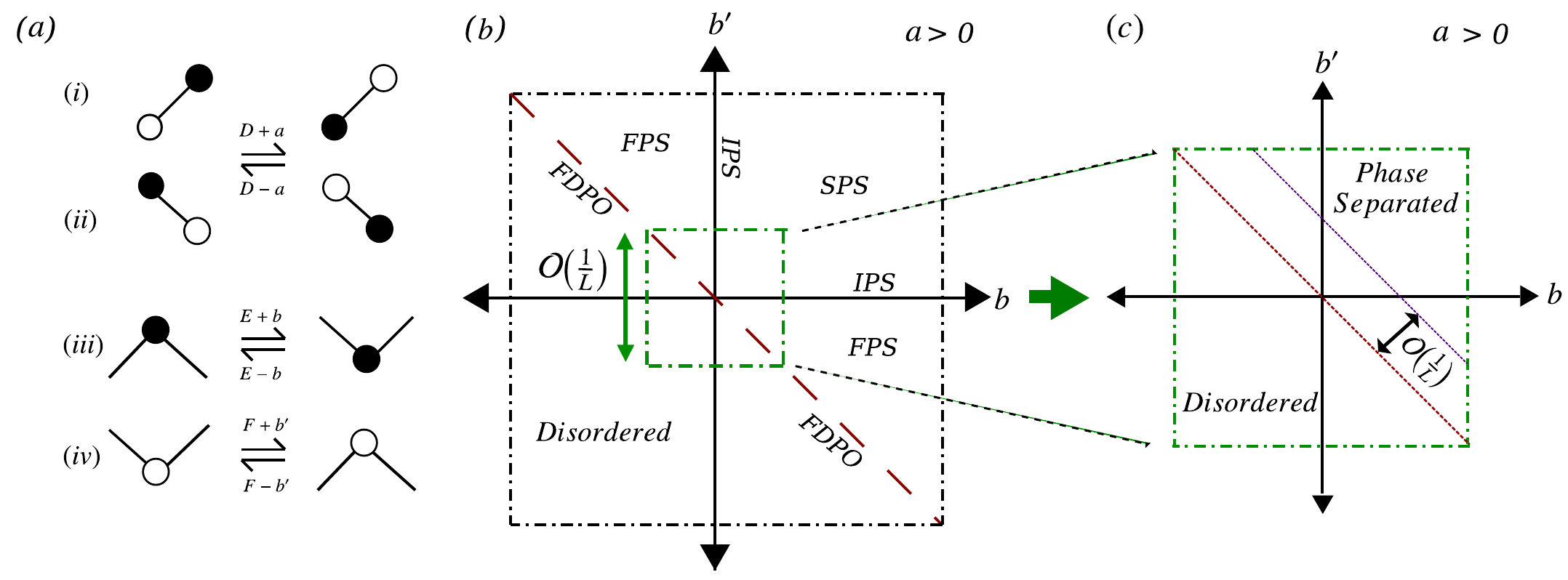}
\caption{
(a) A schematic representation of the local update rules in the Light-Heavy (LH) model. (b) The $2D$ phase diagram of the uLH model shows three ordered states: strong phase separation (SPS), infinitesimal current with phase separation (IPS), and finite current with phase separation (FPS), as well as Fluctuation-Dominated Phase Ordering along the $b+b^{\prime}=0$ plane and a disordered phase.
(c) In the $2D$ phase diagram of the sLH model, we observe an ordered phase that exhibits normal phase separation, a disordered phase, and a critical plane separating the two. This model exhibits a continuous phase transition from the disordered to the ordered phase, with theoretical predictions~\cite{Chakraborty_2017} suggesting $\sqrt{a^{2}_{c}+b^{2}_{c}+b^{\prime{2}}_{c}} \sim 1/L$ along specific loci in the parameter space along which detailed balance is obeyed. Numerical studies indicate shifts in the critical points from $b+b^{\prime}=0$ plane while retaining the planar structure of the critical surface throughout the parameter space.
}
\label{pic2}
\end{figure}

The model exhibits coupled dynamics. The local particle configurations influence the tilt dynamics and vice versa. The system can therefore be thought of as a collection of particles on an uneven surface, which causes the particles to move while also being pushed down or pulled up by the particle between two successive tilts. The exact rates of local updates are dictated by the parameters $a$, $b$, and $b^{\prime}$ and are given in Figure~\ref{pic2}a. Rules ($i$) and ($ii$) explain how the surface influences particles, resulting in particle exchanges between neighboring sites at rates $D \pm a$ (where $D$ is the diffusive rate and $a$ is the bias rate). In contrast, rules ($iii$) and ($iv$) describe particle interactions with the surface at rates $E \pm b$ and $F \pm b^{\prime}$ (where $E,~F$ are the diffusive rates and $b, b^{\prime}$ are the bias rates).
We categorize the regime of unscaled bias rates ($a, b, b^{\prime} \sim \mathcal{O}(1)$) as the unscaled Light-Heavy (uLH) model, while scaled bias rates ($a, b, b^{\prime} \sim \mathcal{O}(1/L)$) constitute the scaled Light-Heavy (sLH) model. Monte Carlo simulations of this model are performed by randomly choosing a particle and a tilt in every microscopic time step ($t$, $t$ + $\Delta t$), and allowing them to evolve based on the transition probabilities set by the tilts or particles that are on either side. We allow for as many microscopic steps as the size of the system so that every particle and tilt are on average updated at each time step.

The Light-Heavy model has been studied in both the unscaled (uLH) and scaled (sLH) regimes. In the former, the bias rates are of $\mathcal{O}(1)$ while in the latter, they are of $\mathcal{O}(1/L)$.
The coupled dynamics in the system leads to multiple steady states that occupy different regions of the 3D parametric space constructed of $a$, $b$, and $b^{\prime}$ in uLH~\cite{Chakraborty_2016}. Figure~\ref{pic2}b shows a 2D subspace of this 3D region, where $a$ is set to be greater than zero, or in other words, the particles move as if under the effect of the gravitational field, with heavier ones wanting to move downhill, and the lighter ones - uphill.
The consequent steady states in the system include three types of ordered states, a disordered state, and a special state characterized by Fluctuation-Dominated Phase Ordering (FDPO) which separates the ordered and disordered phases. Illustrations of these different phases can be found in~\cite{Mahapatra2020Light}. All the ordered states show strong clustering of like particles, i.e., they form a single macroscopic cluster which is capped by an interfacial region of the size of a few lattice spacings, and differ from each other in the extent of ordering in the tilts. 

The first of the ordered states is the strong phase separation {\it (SPS)}, where particles and tilts, are both phase separated. Here, the surface is made of sharp hills and valleys, with heavy particles clustered near valleys and lighter particles near hills. We observe such a state when $b$ and $b^{\prime}$ are
positive, heavier particles push hills down, and lighter particles pull valleys up. 

Alternatively, if we set either $b$ or $ b^{\prime}$ at zero while allowing the other to take on finite positive values, the resulting state is referred to as an infinitesimal current with phase separation {\it (IPS)}.
When $b$ (or $b^{\prime}$) is zero while $b^{\prime}$~(or $b$) has a positive finite value, the tilts remain unaffected by the heavy (or light) particles, resulting in a perfect hill (or valley) while the valley (or hill) becomes more parabolic in shape. The valley (or hill) then acts as a reservoir for the tilt current, which is infinitesimal in magnitude and flows through the system~\cite{Chakraborty_2016}. Despite changes to the tilt profile, the particles remain strongly phase separated in this phase.

The last of the ordered states is the one with finite current with phase separation {\it (FPS)}, where $b+b^{\prime}> 0$, and $b$, or $b^{\prime}$ $< 0$. Both particle species push the surface down in this region, albeit to different extents. The resultant structure shows an imperfect ordering corresponding to the positive parameter ($b$ or $b^{\prime}$) and a disorder for the negative parameter ($b^{\prime}$ or $b$). The system generates finite tilt currents, which gives the phase its name~\cite{Chakraborty_2016}. 

In addition, we have Fluctuation-Dominated Phase Ordering {\it (FDPO)} at the order-disorder boundary ($b$ + $b^{\prime} = 0$), which shows macroscopic clusters of the order of system size, which are continuously reorganizing, indicating large fluctuations~\cite{barma2023fluctuationdominated, das2000particles, das2001fluctuation, Kapri_2016, Barma_2019, Chatterjee_2006, Das_2023}. Finally, the disordered phase with $b + b^{\prime}< 0$ lacks order in the particles and tilts as the particles drive the surface in a direction that opposes the path of particle drift that the surface induces.
\newline
All these diverse states arise only within the parameter space of unscaled rates (uLH), i.e., when the bias rates are comparable to the diffusive rates. If we enter the scaled regime (sLH), i.e., the domain of scaled rates,
\begin{equation}
\begin{split}
    a = \alpha/L, \\
    b = \beta/L,\\
    b^{\prime} = \beta^{\prime}/L,
\end{split}
\end{equation}
where $a, b, b^{\prime} \sim \mathcal{O}(1/L)$, then, we find that most of these phases cease to exist, leaving an ordered and disordered phase separated by the same critical line as before ($b+b^{\prime} = 0$). See Section~\ref{sec: phase transition} for details on this change in the phase diagram.

An earlier study~\cite{Chatterjee_2006} used a mean-field approach to conjecture the following noiseless hydrodynamics for the unscaled model

\begin{equation}
 \frac{\partial \rho}{\partial t}    = D \frac{\partial^{2} \rho}{\partial x^{2}} +\frac{\partial}{\partial x} \bigg[2\alpha\rho(1-\rho)(1-2m)\bigg],
 \nonumber
\end{equation}
\begin{equation}
 \frac{\partial m}{\partial t}    = D \frac{\partial^{2} m}{\partial x^{2}} +\frac{\partial}{\partial x} \bigg[m(1-m)(2\rho(\beta+\beta^{\prime})-2\beta^{\prime})\bigg].
 \label{noiseless}
\end{equation}
In this context, $\rho$ represents the coarse-grained density of heavy particles, while $m$ denotes the density of upward tilts. In fact, these expressions turn out to be exact as seen by comparing the results obtained from an action formulation discussed in Section~\ref{sec:Action-LH}. This formalism enables us to find the noise correlations exactly in the scaled model.

\section{Action formalism and fluctuating hydrodynamics}
\label{sec:Action-LH}
Recent studies have used fluctuating hydrodynamics to establish an analytical description of the correlation function for similar systems~\cite{agranov2021exact}. In this section, we apply this method to the LH model.
Macroscopic Fluctuation Theory (MFT) is a recent innovation that has been successfully used in such efforts. The formalism works by adding an appropriate noise term to a system whose exact hydrodynamic equation can be derived from a microscopic picture. We do this to incorporate the fluctuating noisy dynamics which sets in when we investigate properties of finite systems instead of infinite systems. Agranov~{\it et al.}~\cite{agranov2021exact} recently used this framework to examine the correlation length of active lattice gases in regions of the phase diagram without phase separation. In this work, we do the same for the LH model. By developing a suitable fluctuating hydrodynamic description of the system, we derive two-point correlation functions in the disordered phase and analyze the critical behavior of the model.

The first step in this endeavor is to derive the fluctuating hydrodynamic equations. To do this, we use the Lef\`evre
-Biroli method~\cite{Andreanov_2006, Lefevre_2007}, which can be shown to be exactly the same as the Doi-Peliti formulation~\cite{peliti1985path, Doi_1976, Doi_1976_2}. Using this technique~\cite{ Thompson_2011, wijland_2001}, we can find the exact fluctuating hydrodynamics for the Light-Heavy model. Here, we will limit ourselves to outlining the procedure and stating the final expression, while reserving the details for~\ref{sec:appendix b}. 
This method uses a path integral formalism to derive the fluctuating hydrodynamics of a system starting from its local update rules. Given a system that undergoes dynamics with a predefined initial and final state, there are many paths within the configuration space that it may take. However, not all paths are equally likely. The local update rules of the system determine how probable the different paths are. Using this information, we can establish a path-integral formalism for the probability of observing a given trajectory of particles and tilts.

Consider a lattice of $L$ sites, each separated by a distance of $\epsilon$ at a discretized time $t_{j}$ with $j \in {1,2,\hdots N}$. To describe the various changes that may occur in the system with time, we first define the following variables,\\
\par
$\eta^{\sigma}_{i}(t_{j}) = 
        \begin{array}{cc}
         \Biggl\{ &
          \begin{matrix}
           0 & \text{Light}  &\circ\\
           1  & \text{Heavy}  &\bullet    
          \end{matrix}
        \end{array}\hspace{3cm}
\eta^{\tau}_{i}(t_{j}) = 
        \begin{array}{cc}
         \Biggl\{ &
          \begin{matrix}
           0    &\text{Down} &\setminus \\ 
           1    &\text{Up}  &/
          \end{matrix}
        \end{array}$
\vspace{5mm}\\
A trajectory of the system is then ultimately determined by the set ${\eta}$ containing all $\eta^{\sigma,\tau}_{i}(t_{j})$
\begin{equation}
\eta = \biggl\{ \eta^{\sigma,\tau}_{i}(t_{j}) \hspace{1cm} \text{for}  \hspace{5mm} i \in {1,2,\hdots L},\hspace{5mm} j \in {1,2,\hdots N}.
\end{equation}
Specifically, we can also say that at time $t_{j}$, the entire system of particles and tilts are described by the set ${\eta^{j}}$ defined as
\begin{equation}
\eta^{j} = \biggl\{ \eta^{\sigma,\tau}_{i}(t_{j}) \hspace{1cm} \text{for} \hspace{5mm} i \in {1,2,\hdots L}.
\end{equation}
In a time $\delta = t_{j+1}-t_{j}$, the system undergoes one of the eight processes described in Figure~\ref{pic2}a. Here, we set the diffusive rate of the particle and tilt evolution to $D$. Another possibility is that the system may stay as it is, with no change occurring in its previous state. In either way, we need to keep track of the changes or lack thereof in the system over a time interval $d\tau$. To this end, we define the quantity $J^{\sigma,\tau}_{i}(t_{j})$ as the variation in the state of particle or tilt at site $i$, between time $t_{j}$ and $t_{j+1}$
\begin{equation}
\begin{split}
J^{\sigma}_{i}(t_{j}) = \eta^{\sigma}_{i}(t_{j+1}) - \eta^{\sigma}_{i}(t_{j}),\\
J^{\tau}_{i}(t_{j}) = \eta^{\tau}_{i}(t_{j+1}) - \eta^{\tau}_{i}(t_{j}).
\end{split}
\end{equation}
Since at most one particle or tilt can undergo some change in a unit time $\delta$, each $J^{\sigma, \tau}_{i}(t_{j})$ takes values in ${1,0,-1}$ only and at most two of them can be non-zero at a particular time $t_{j}$.
We can therefore introduce another set ${J}$ which contains all $J^{\sigma, \tau}_{i}(t_{j})$,
\begin{equation}
J = \biggl\{ J^{\sigma, \tau}_{i}(t_{j}) \hspace{1cm} \text{for}  \hspace{5mm} i \in {1,2,\hdots L},\hspace{5mm} j \in {1,2,\hdots N}.
\end{equation}
As with $\eta$, we can also combine all $J^{\sigma,\tau}_{i}(t_{j})$ at time $t_{j}$, to define the set ${J^{j}}$ as
\begin{equation}
J^{j} = \biggl\{ J^{\sigma, \tau}_{i}(t_{j}) \hspace{1cm} \text{for} \hspace{5mm} i \in {1,2,\hdots L}.
\end{equation}
At this point, we can establish a path integral for probability $P(\{\eta\})$ to observe a given trajectory $\{\eta\}$ of particles and tilts.
\begin{equation}
P(\{\eta\}) = \Bigg \langle \prod_{i=1}^{L} \prod_{j=1}^{N} \delta \big(\eta^{\sigma}_{i}(t_{j+1}) - \eta^{\sigma}_{i}(t_{j}) - J^{\sigma}_{i}(t_{j})\big) \delta \big(\eta^{\tau}_{i}(t_{j+1}) - \eta^{\tau}_{i}(t_{j})-J^{\tau}_{i}(t_{j})\big) \Bigg \rangle _{\{J\}}.
\end{equation}
We can express this equation as an integral by using the integral form of the delta function, i.e.,
\begin{equation}
\delta(x) = \int d \hat{x} e^{-i x \hat{x}}.
\end{equation}
Then, by introducing fields conjugate to $\eta^{\sigma}_{i}(t_{j})$ and $\eta^{\tau}_{i}(t_{j})$, given by $\hat{\rho}_{i}(t_{j})$ and $\hat{m}_{i}(t_{j})$, we obtain
\begin{equation}
\begin{split}
P(\{\eta\}) = \int \prod_{J=1}^{N} \Bigg[ \prod_{i=1}^{L} \Bigl[ d\hat{\rho}_{i}(t_{j}) d\hat{m}_{i}(t_{j})  e^{-\hat{\rho}_{i}(t_{j})\big[\eta^{\sigma}_{i}(t_{j+1}) - \eta^{\sigma}_{i}(t_{j})\big]- \hat{m}_{i}(t_{j}) \big[\eta^{\tau}_{i}(t_{j+1}) - \eta^{\tau}_{i}(t_{j})\big]} \Bigr]\\
 \Bigg \langle \prod_{i=1}^{L} \Bigl[e^{\hat{\rho}_{i}(t_{j}) J^{\sigma}_{i}(t_{j})+\hat{m}_{i}(t_{j}) J^{\tau}_{i}(t_{j})}\Bigr]\Bigg \rangle _{\{J\}} \Bigg].
 \label{eq:probdistri}
\end{split}
\end{equation}

The local update rules are then incorporated into Eq.~\eqref{eq:probdistri} to appropriately weigh the different possible trajectories. 

The occupation variables are coarse-grained using boxes of width $\Delta$, where $\Delta \sim L^{\delta}$ with $\delta <1$, creating local densities that facilitate our transition to continuum space
\begin{align}
    \rho(x,t) &= \lim_{\epsilon\rightarrow0} \Bigg(\frac{1}{\Delta} \sum\limits_{x_i = x_i}^{x_i + \Delta} \eta^{\sigma}(x_{i},t_{j})\Bigg), \\
    m(x,t) &= \lim_{\epsilon\rightarrow0} \Bigg(\frac{1}{\Delta} \sum\limits_{x_i = x_{i+\frac{1}{2}}}^{x_{i+\frac{1}{2}} + \Delta} \eta^{\tau}(x_{i},t_{j})\Bigg).
\end{align}

Using this coarse-graining procedure, we arrive at a continuum description
\begin{equation}
P(\{\eta\}) = \int \mathcal{D}[\hat{\rho},\hat{m}]e^{L\int \int dx dt \mathcal{S}},
\end{equation}
where the continuum action is,
\begin{equation}
\begin{split}
\mathcal{S} = \dot{\rho} \hat{\rho} + \dot{m} \hat{m} +  D \hat{\rho}_{x} \rho_{x} + D \hat{m}_{x} m_{x} +
D\rho (1-\rho)\hat{\rho}_{x}^{2} + Dm (1-m)\hat{m}_{x}^{2} \\+ 2 \alpha \rho (1-\rho) (1-2m)\hat{\rho}_{x} + 
2 m(1-m)\big((\beta+\beta^{\prime})\rho- \beta^{\prime}\big) \hat{m}_{x}.
\end{split}
\end{equation}

\begin{figure}[t!]
\centering
\includegraphics[width=0.75\linewidth] {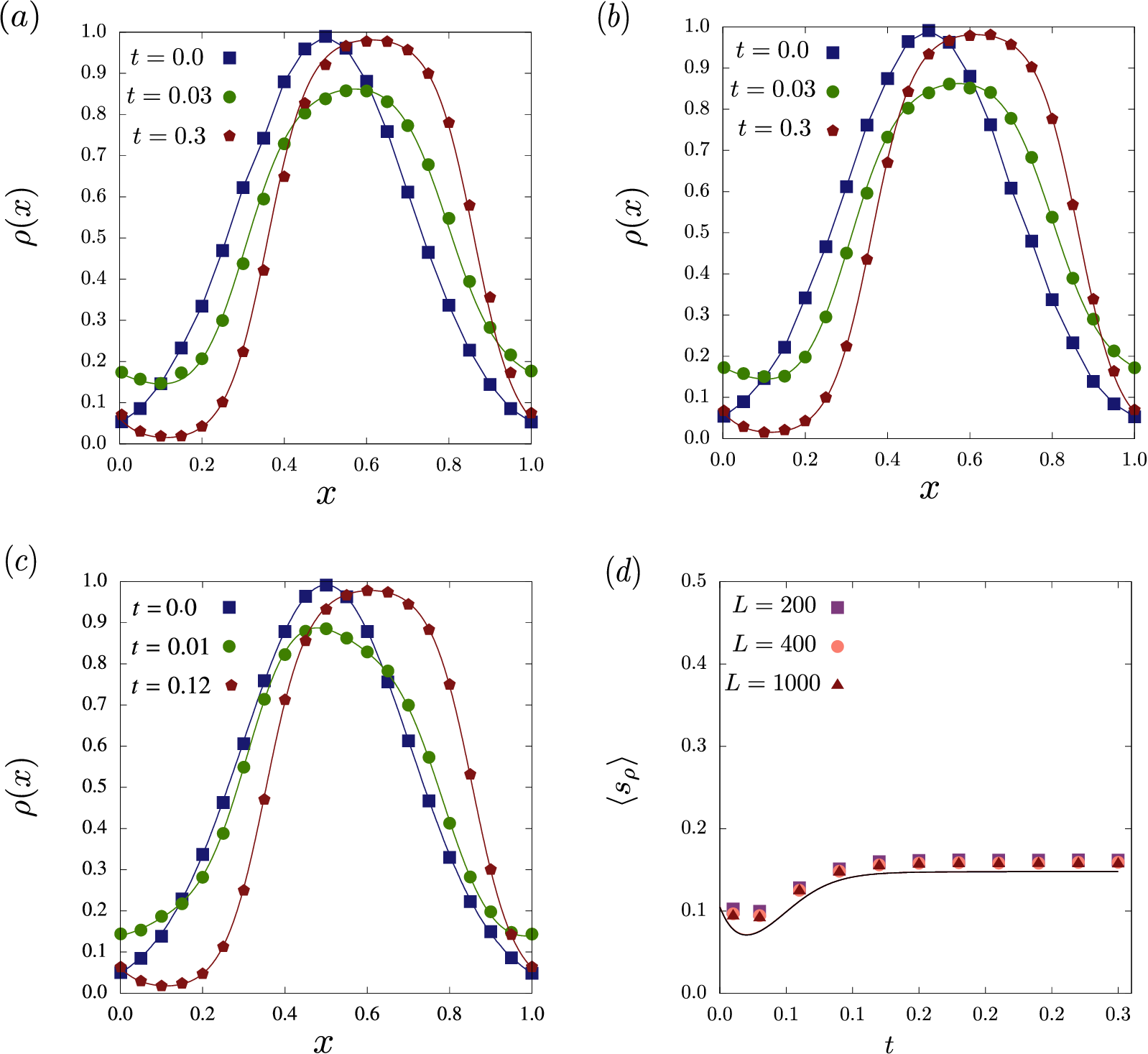}
\caption{Comparison of hydrodynamic predictions with numerical simulations. In panels ($a$), ($b$), and ($c$), we illustrate the evolution of an initial Gaussian particle density profile for systems with sizes $L = 200, 400$ and $1000$. Panel ($d$) displays the corresponding order parameter, defined as $\langle S_{\rho} \rangle = \frac{1}{L} \sum_{i} (\rho_{i} - \rho_{0})^{2}$ for all three systems. In all the panels, the symbols represent the predictions from simulations, while the lines reflect the theoretical results derived from noiseless hydrodynamics as given by Eq.~\eqref{eq:noiseless}. The parameters used are $\alpha = 5, \beta = 7$ and $\beta' = 9$. The data presented has been averaged over $1000$ histories.}
\label{hyd_mag}
\end{figure}

The Euler-Lagrange equations of motion are then given by
\begin{equation}
 \frac{\partial \rho}{\partial t}    = D \frac{\partial^{2} \rho}{\partial x^{2}} +\frac{\partial}{\partial x} \bigg[2\alpha\rho(1-\rho)(1-2m)\bigg] + \frac{\partial [2D \rho (1-\rho)\hat{\rho}_{x}]}{\partial x},
 \label{eq:partdiffaaction3}
 \nonumber
\end{equation}
\begin{equation}
\frac{\partial m}{\partial t}  = D \frac{\partial^{2} m}{\partial x^{2}} +\frac{\partial}{\partial x} \bigg[m(1-m)(2\rho(\beta+\beta^{\prime})-2\beta^{\prime})\bigg] +  \frac{\partial [2D m (1-m)\hat{m}_{x}]}{\partial x}, \label{eq:partdiffaaction}
\end{equation}
with the noiseless part given by the coupled equations,
\begin{equation}
 \frac{\partial \rho}{\partial t}    = D \frac{\partial^{2} \rho}{\partial x^{2}} +\frac{\partial}{\partial x} \bigg[2\alpha\rho(1-\rho)(1-2m)\bigg],
 \nonumber
\end{equation}
\begin{equation}
\frac{\partial m}{\partial t}    = D \frac{\partial^{2} m}{\partial x^{2}} +\frac{\partial}{\partial x} \bigg[m(1-m)(2\rho(\beta+\beta^{\prime})-2\beta^{\prime})\bigg].
\label{eq:noiseless}
\end{equation}
Interestingly, the noiseless equations are the same as those stated in Eq.~\eqref{noiseless}, which were derived in~\cite{Chatterjee_2006} using a phenomenological approach. We show the validity of the above noiseless hydrodynamic equations in Figure~\ref{hyd_mag} by comparing them with the predictions from Monte Carlo simulations. 

To obtain the fluctuating hydrodynamic equations for the LH model, we need to determine the form of the noise correlations. We do this by coupling a noise term to the hydrodynamic equations. This suggested form is then used to formulate a path integral, which is then used to calculate an action. Then, a simple comparison with Eq.~\eqref{eq:partdiffaaction} reveals the fluctuating hydrodynamic equations. These turn out to be,
\begin{equation}
 \frac{\partial \rho}{\partial t}    = D \frac{\partial^{2} \rho}{\partial x^{2}} +\frac{\partial}{\partial x} \bigg[2\alpha\rho(1-\rho)(1-2m)\bigg] + \frac{\partial \eta_{\rho}}{\partial x},
 \label{hyd1}
 \nonumber
\end{equation}
\begin{equation}
 \frac{\partial m}{\partial t}    = D \frac{\partial^{2} m}{\partial x^{2}} +\frac{\partial}{\partial x} \bigg[m(1-m)(2\rho(\beta+\beta^{\prime})-2\beta^{\prime})\bigg] + \frac{\partial \eta_{m}}{\partial x},
 \label{hyd2}
\end{equation}
with the noise correlations given by
\begin{eqnarray}
 \langle \eta_{\rho}(x,t)  \eta_{\rho}(x^{\prime},t^{\prime})\rangle = \frac{2 D \rho (1-\rho)}{L} \delta (x-x^{\prime}) \delta (t-t^{\prime}),
 \nonumber\\
 \langle \eta_{m}(x,t)  \eta_{m}(x^{\prime},t^{\prime})\rangle = \frac{2 D m (1-m)}{L} \delta (x-x^{\prime}) \delta (t-t^{\prime}),
 \nonumber\\
 \langle \eta_{\rho}(x,t)  \eta_{m}(x^{\prime},t^{\prime})\rangle = 0.\nonumber\\
\end{eqnarray}

At this point, it is useful to discuss a related model that has been studied using similar techniques in the literature, namely the active lattice gas~\cite{Kourbane_2018,agranov2021exact,jose2023current}. 
This model, defined in a one-dimensional ring lattice, comprises $L$ sites and accommodates $N$ particles, resulting in a density $\rho_{0} = N/L$.
Each particle is encoded with a preferential drift direction to introduce activity into the system. Doing so classifies the entire set of particles into $+$ and $-$ species. This $\pm$ nature attributed to a particle is not permanent and changes with time as part of the dynamics of the system. The entirety of the dynamics of the active lattice gas is defined in terms of three independent processes.
\begin{itemize}
\item Diffusion: a pair of adjacent sites interchange particles with rate D.
\item Drift: a $+$ ($-$) particle hops to the right (left) neighbor site with a rate $\lambda/L$, if the site is empty.
\item Tumble: a $+$ ($-$) particle tumbles into a $-$ ($+$) one with a rate $\gamma/L^{2}$.
\end{itemize}
The rates are scaled with $L$ to ensure that all processes occur on the diffusive time scale in the hydrodynamic limit.
The noiseless hydrodynamic description of this model has been derived in~\cite{Kourbane_2018}. To this, Gaussian noise terms were added by Agranov {\it et al.}~\cite{agranov2021exact} to capture the fluctuations to the profile predicted by this description in a finite lattice. They did so by superposing known results for the ABC model with a reaction term to account for the tumbling. Alternatively, one can arrive at the same expression using a path integral technique~\cite{peliti1985path}. The fluctuating hydrodynamic equations for the system were determined to be,
\begin{eqnarray}
\partial_t \rho&=& D \partial^2_{x}\rho-\lambda\partial_x[m(1-\rho)]+\partial_x\eta_\rho\label{eq:rho},
\nonumber\\
\partial_t m&=& D \partial^2_{x}m-\lambda\partial_x[\rho(1-\rho)]-2\gamma m+\partial_x\eta_m+\eta_K \,. \label{eq:m}
\end{eqnarray}
The terms $\eta_{\rho}, \eta_{m}$, and $\eta_{K}$ in Eq.~\eqref{eq:m} represent zero-mean Gaussian white noise, characterized by the following correlations
\begin{eqnarray}
\langle\eta_{\rho} (x,t)\eta_{\rho} (x',t')\rangle&=&\frac{2 D \rho (1-\rho)}{L}\delta (x-x')\delta (t-t'),\nonumber\\
\langle\eta_{\rho} (x,t)\eta_{m} (x',t')\rangle&=&\frac{2 D m (1-\rho)}{L}\delta (x-x')\delta (t-t'), \nonumber
\label{covrho}\\
\langle\eta_{m} (x,t)\eta_{m} (x',t')\rangle&=&\frac{2 D (\rho-m^2)}{L}\delta (x-x')\delta (t-t'),\label{covm}\\
\langle\eta_K (x,t)\eta_K (x',t')\rangle&=&\frac{4 \gamma \rho}{L}\delta (x-x')\delta (t-t')\nonumber,
\label{covk} \\
\langle\eta_{m} (x,t)\eta_{K} (x',t')\rangle&=&\langle\eta_{\rho} (x,t)\eta_{K} (x',t')\rangle= 0 \,. \nonumber
\label{eq:cov0}
\end{eqnarray}

These equations share similarities with those we previously identified for the LH model, but there are three primary differences. First, although both models describe the hydrodynamics of two coupled fields, the nature of the drift terms are different.
Second, in contrast to the LH model, the ALG model includes a noise term, denoted
$\eta_{k}$, which arises from tumbling events and follows Poissonian statistics. Finally, the nature of the noise correlation differs between the two models. In the case of the ALG model, the correlations $\eta_m-\eta_m$ involve both fields $\rho$ and $m$, as demonstrated in Eq.~\eqref{covm}. However, in the Light-Heavy model, the correlation between the noise terms is influenced solely by the density of the same species.

Yet another model that displays similar hydrodynamics is the ABC model~\cite{Clincy_2003, Bodineau_2011}. As mentioned earlier, this model served as a template for deriving the fluctuating hydrodynamics for the ALG model. The ABC model comprises three species of particles with hardcore interaction, named $A$, $B$, and $C$, with densities $\rho_{A}$, $\rho_{B}$ and $\rho_{C}$. The dynamics of the ABC model are given by \\

\hspace{5cm}\ce{AB <=>[\text{q}][\text{1}] BA},\\

\hspace{5cm}\ce{BC <=>[\text{q}][\text{1}] CB},\\

\hspace{5cm}\ce{CA <=>[\text{q}][\text{1}] AC}.\\
Here, by scaling $q$ as $q = e^{-\beta/N}$, we transition to the scaled ABC model, which obeys the following hydrodynamics:
\begin{eqnarray}
\nonumber
\partial_t \rho_{A}&=& D \partial^2_{x}\rho_{A}+\beta\partial_x[\rho_{A}(\rho_{A} + 2 \rho_{ B}-1)]+\partial_x\eta_{\rho_{A}}\label{eq:abc1},\\
\partial_t \rho_{B}&=& D \partial^2_{x}\rho_{B}+\beta\partial_x[\rho_{B}(1-2\rho_{A} - \rho_{ B})]+\partial_x\eta_{\rho_{B}}. \label{eq:abc2}
\end{eqnarray}
Here, $\eta_{\rho_{A}}$ and $\eta_{\rho_{B}}$ are zero-mean Gaussian white noises with the following correlations
\begin{eqnarray}
\nonumber
\langle\eta_{\rho_{A}} (x,t)\eta_{\rho_{A}} (x',t')\rangle&=&2\rho_{A} (1-\rho_{A})\,\delta (x-x')\delta (t-t'),\\
\nonumber
\langle\eta_{\rho_{B}} (x,t)\eta_{\rho_{B}} (x',t')\rangle&=&2\rho_{B} (1-\rho_{B})\,\delta (x-x')\delta (t-t'),
\label{abc3}\\
\langle\eta_{\rho_{A}} (x,t)\eta_{\rho_{B}} (x',t')\rangle&=&-2 \rho_{A} \rho_{B}\delta (x-x')\delta (t-t').\label{abc4}
\end{eqnarray}
While the general structure of this hydrodynamics is the same as the LH model, one ought to note that the mixed correlations are non-zero here.

\section{Derivation of correlation functions}
\label{sec:derivation-two-point}
In this section, we use the fluctuating hydrodynamics of the sLH model, which we derived in the previous section, to calculate both the static and dynamic two-point correlation functions in the disordered phase of the model.
\subsection{Static correlation functions}
\label{sec:static}

We begin by computing the static (steady state) correlation functions between the density fields in the sLH model, defined as
\begin{eqnarray}
\nonumber
\mathcal{C}_{2}^{\rho}(x,x^{\prime}) = \langle \delta \rho(x)\delta \rho(x^{\prime})\rangle, \\
\nonumber
\mathcal{C}_{2}^{m}(x,x^{\prime}) = \langle \delta m(x)\delta m(x^{\prime})\rangle, \\
\mathcal{C}_{2}^{\rho,m}(x,x^{\prime}) = \langle \delta \rho(x)\delta 
 m(x^{\prime})\rangle.
\end{eqnarray}
Here, $\delta\rho(x)$ and $\delta m(x)$ are local fluctuations in the density of particles and tilts about their global average in the steady-state. We introduce the parameter $\Delta$ in terms of the scaled rates $\beta$ and $\beta^{\prime}$, which describes the distance to the critical line.

\begin{equation}
\beta = -\beta^{\prime} - \Delta.
\end{equation}
Then,
\begin{equation}
 \frac{\partial m}{\partial t}    = D \frac{\partial^{2} m}{\partial x^{2}} +\frac{\partial}{\partial x} \bigg[m(1-m)(-2\rho\Delta-2\beta^{\prime})\bigg] + \frac{\partial \eta_{m}}{\partial x}.
\end{equation}
In the hydrodynamic limit, we can obtain the solutions for the two-point correlation function by linearising these equations for small fluctuations $\rho(x,t) = \rho_{0} + \delta  \rho(x,t)/\sqrt{L}$ and $m(x,t) = m_{0} + \delta m(x,t)/\sqrt{L}$. This gives,
 \begin{equation}
 \frac{\partial \delta \rho}{\partial t}    = D \frac{\partial^{2} \delta \rho}{\partial x^{2}}   -4\alpha\rho_{0}(1-\rho_{0})\frac{\partial\delta m}{\partial x} + 2\alpha(1-2\rho_{0})(1-2m_{0})\frac{\partial\delta \rho}{\partial x} +\frac{\partial \eta_{\rho}}{\partial x},
 \nonumber
\end{equation}
\begin{equation}
 \frac{\partial \delta m}{\partial t}    = D \frac{\partial^{2} \delta m}{\partial x^{2}} +2\Delta m_{0}(1-m_{0})\frac{\partial\delta \rho}{\partial x} + (1-2m_{0})(-2\rho_{0}\Delta + \beta^{\prime})\frac{\partial\delta m}{\partial x} + \frac{\partial \eta_{m}}{\partial x}.
\end{equation}
We can further simplify the expression by taking into account the fact that we are looking at systems with no net tilt on the surface, i.e., $m_{0} = \frac{1}{2}$. Then
 \begin{equation}
 \frac{\partial \delta \rho}{\partial t}    = D \frac{\partial^{2} \delta \rho}{\partial x^{2}}   -4\alpha\rho_{0}(1-\rho_{0})\frac{\partial\delta m}{\partial x}+\frac{\partial \eta_{\rho}}{\partial x},
 \label{eq:lin1}
\end{equation}
\begin{equation}
 \frac{\partial \delta m}{\partial t}    = D \frac{\partial^{2} \delta m}{\partial x^{2}} -\frac{\Delta}{2} \frac{\partial \delta \rho}{\partial x}+ \frac{\partial \eta_{m}}{\partial x}.
  \label{eq:lin2}
\end{equation}
Since Eqs.~\eqref{eq:lin1} and \eqref{eq:lin2} are coupled in a nontrivial manner, solving them in real space is challenging. Therefore, we analyze the above equations in Fourier space. Using continuum Fourier transforms given by,
\begin{equation}
	\delta\tilde{m}\left(k,t \right)= \frac{1}{2 \pi} \int dx~ e^{-i kx }\delta m\left(x,t\right)
	\quad;\quad
	\delta\tilde{\rho}\left(k,t \right)= \frac{1}{2 \pi} \int dx~ e^{-ikx}\delta\rho\left(x,t\right),
 \label{eq:FourierTransform2}
\end{equation}
we obtain the set of linear equations
\begin{equation}
 \partial_{t} \begin{pmatrix} \delta \Tilde{\rho}(k,t) \\ \delta \Tilde{m}(k,t) \end{pmatrix} = -\mathcal{M}_{k} \begin{pmatrix} \delta \Tilde{\rho}(k,t) \\ \delta \Tilde{m}(k,t) \end{pmatrix} + \mathcal{R}(k,t),
 \label{eq:lineareqn}
\end{equation}
where 
\begin{equation}
   \mathcal{M}_{k} = \begin{pmatrix} Dk^{2} & iCk\\ iEk & Dk^{2} \end{pmatrix}; \hspace{1.5cm}  \mathcal{R}(k,t) = \begin{pmatrix} Aik\Tilde{\eta_{1}}(k,t)\\ Bik\Tilde{\eta_{2}}(k,t) \end{pmatrix}.
   \label{eq:matrix}
\end{equation}
Here, $A=\sqrt{\frac{2D\rho_{0}(1-\rho_{0})}{L}}$, $B=\sqrt{\frac{2Dm_{0}(1-m_{0})}{L}}$, $C=4\alpha\rho_{0}(1-\rho_{0})$ and $E = \frac{\Delta}{2}$, and $\Tilde{\eta}_{1,2}$ are Gaussian white noises with variance
\begin{equation}
\langle \Tilde{\eta}_{i}(k,t)  \Tilde{\eta}_{j}(k^{\prime},t^{\prime})\rangle =  \frac{\delta_{ij}}{2\pi}\delta (k+k^{\prime}) \delta (t-t^{\prime}).  
\label{eq:noiseftcor}
\end{equation}
\begin{figure}[t!]
\centering
\includegraphics[width=1.0\linewidth] {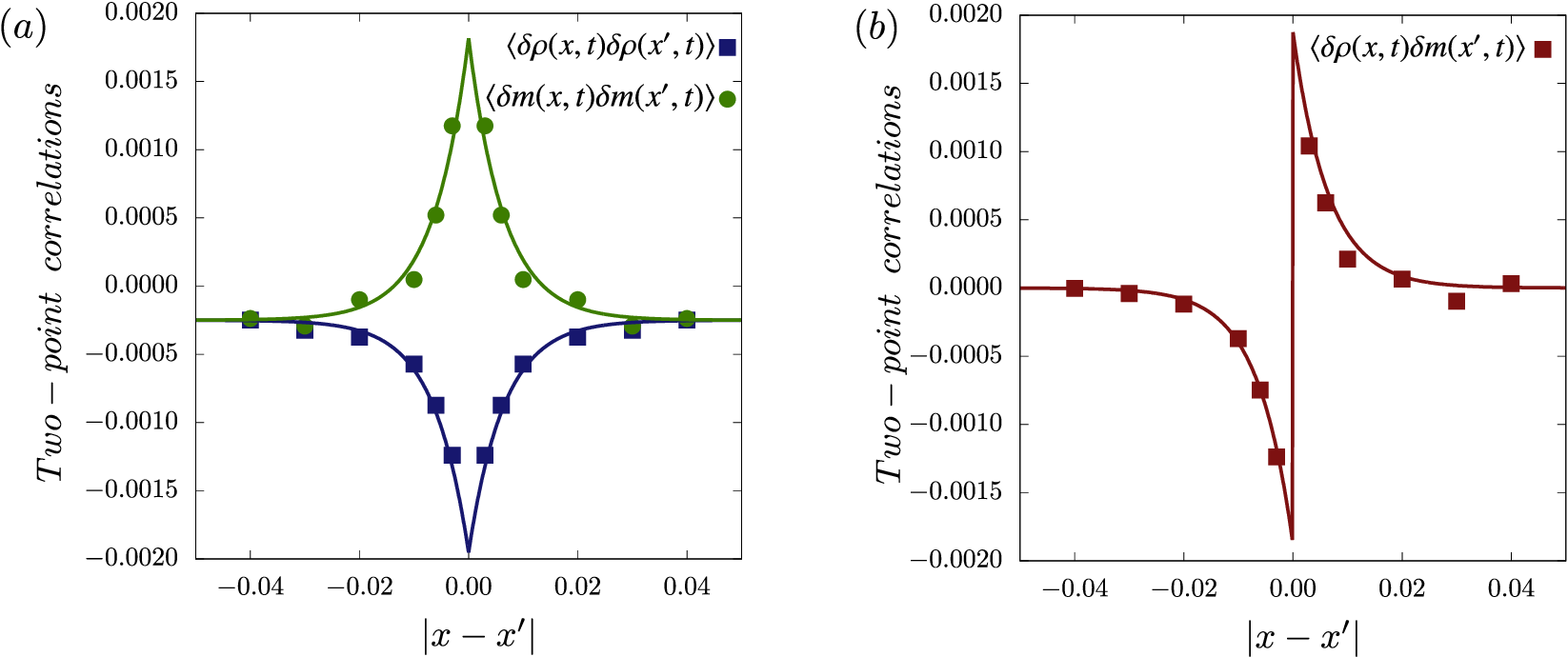}
\caption{Comparison of linearized static density correlations with numerical simulations. The correlation functions for fluctuations in ($a$) particle-particle and tilt-tilt density fields, and ($b$) particle-tilt density field, for a system of size 1000 obtained from performing Monte Carlo simulations (points) plotted against the theoretical result given in Eqs.~\eqref{eq:res1},~\eqref{eq:res2} and~\eqref{eq:res3}  (solid curves). The parameter values used are $\alpha = 70$, $\beta = -90$, and $\beta' = -80$. The simulation data is averaged over 10,000 realizations.}
\label{c1}
\end{figure}
The solution to Eq.~\eqref{eq:matrix} is therefore given by
\begin{equation}
\begin{pmatrix} \delta \Tilde{\rho}(k,t) \\ \delta \Tilde{m}(k,t) \end{pmatrix} =\int_{-\infty}^{t} dt_{1} e^{-\mathcal{M}_{k}(t-t_{1})}\mathcal{R}(k,t_{1}).
\label{lineqn}
\end{equation}

The equal-time correlation function between two Fourier modes of the density field can then be determined from Eq.~\eqref{lineqn} as
\begin{equation}
\mathcal{C}_{2}^{\tilde{\rho}}(k,k^{\prime}) = \langle \delta\tilde{\rho}(k,t) \delta\tilde{\rho}(k^{\prime},t) \rangle.
\end{equation}

We note that while $\delta \rho(k,t)$ explicitly depends on $t$, this time dependence vanishes in the equal-time correlation function $\langle \delta \rho(k,t)\delta \rho(k^{\prime},t)\rangle$. We have
\begin{equation}
\langle \delta \Tilde{\rho}(k',t) \delta \Tilde{\rho}(k,t) \rangle =\int_{-\infty}^{t} dt_{2} \int_{-\infty}^{t} dt_{1} e^{-\mathcal{M}_{k}(t-t_{1})} e^{-\mathcal{M}_{k'}(t-t_{2})} (i A)^{2}\langle \Tilde{\eta}(k,t_{1}) \Tilde{\eta}(k',t_{2}) \rangle.
\label{lineqn}
\end{equation}
The delta-correlation of the noise given in Eq.~\eqref{eq:noiseftcor} simplifies the double integral into a single time integral, which includes a function dependent on $t$ via an exponential factor $e^{-(t - t_1)}$. Thus, integrating $t_{1}$ from $-\infty$ to $t$ removes any remaining $t$-dependence, resulting in a time-independent expression.

\begin{figure}[t!]
\centering
\includegraphics[width=1.0\linewidth] {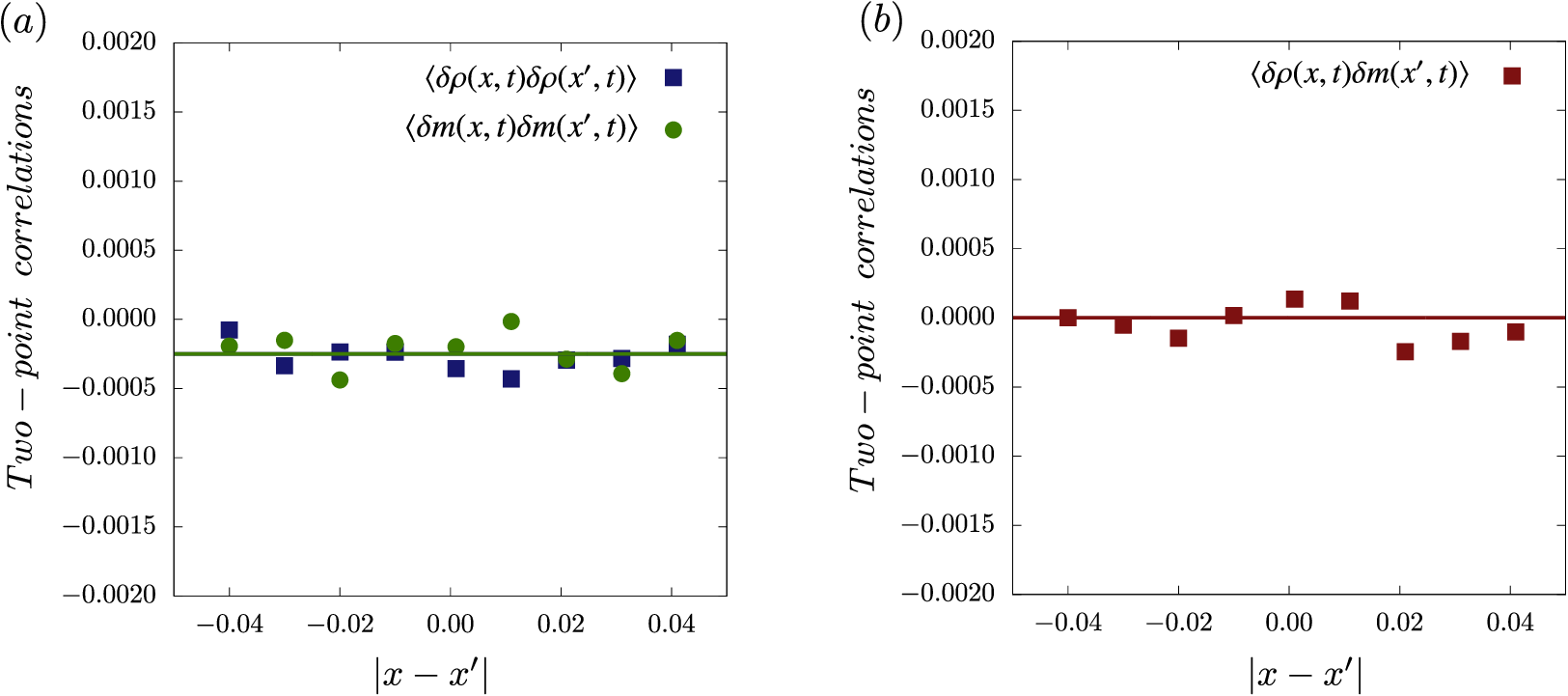}
\caption{The static correlation functions for fluctuations in ($a$) particle-particle, tilt-tilt, and ($b$) particle-tilt density fields for a system of size of $1000$ in a subspace of the phase diagram that obeys bunchwise balance ($2\alpha+\beta+\beta^{\prime} = 0$)~\cite{Mahapatra2020Light}. The parameter values chosen are ($\alpha, \beta, \beta^{\prime}$) $= (70,-90,-50)$. 
Predictions from simulation (points) and theory (lines) given in Eqs.~\eqref{eq:res1}~\eqref{eq:res2} and \eqref{eq:res3} indicate the absence of long-range correlations in this subspace. The Monte Carlo data has been averaged over $3000$ histories.}
\label{res3}
\end{figure}

Conservation of the total number of particles and the total number of tilts ensures that the $k=0$ mode of the densities remains zero at all times. Consequently, the equal-time correlation function at $k=k^{\prime}=0$ will disappear. Considering this, we employ a Dirac delta function $\delta(k)$, with a magnitude $l$ proportional to the separation between modes, to derive steady-state density correlations as
\begin{figure}[t!]
\centering
\includegraphics[width=1.0\linewidth] {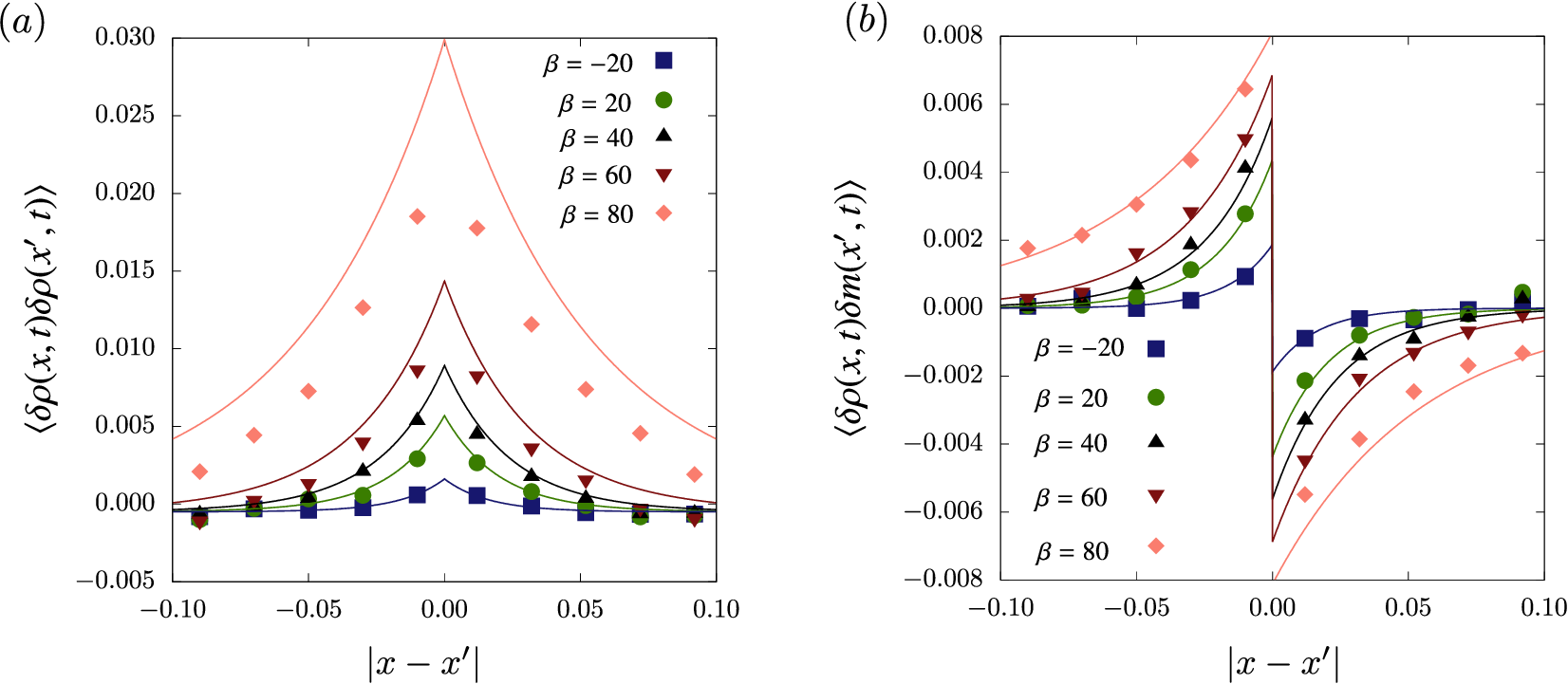}
\caption{Comparison of the static correlation functions for fluctuations in ($a$) particle-particle and ($b$) particle-tilt density field within the homogeneous phase of the phase diagram, on approaching the critical line ($\beta+\beta^{\prime}=0$). Points obtained from simulations indicate a deviation from the solid lines, which represent the theoretical results derived in Eqs.~\eqref{eq:res1} and~\eqref{eq:res3}. This deviation occurs as we get closer to the critical line. The parameter values used for ($\alpha, \beta$) are $70$ and $-90$, respectively while ($\beta^{\prime}$) is sampled across the values $-20, 20, 40, 60$ and $80$. The system size used is $L = 1000$ and the numerical data has been averaged over $2000$ histories.}

\label{res1}
\end{figure}
 
\begin{equation}
\begin{split}
\mathcal{C}^{\tilde{\rho}}_{2}(k,k^{\prime}) = \langle\delta \rho(k)\delta \rho(k^{\prime}) \rangle= \Bigg(\frac{B^2 C^2 +A^2 (CE+2D^2 k^2)}{ 4D (C E + D^2 k^2)}- \frac{B^2C+A^2E}{4DE}l \delta(k)\Bigg)\frac{\delta(k+k^{\prime})}{2\pi},
\label{eq:resft1}
\end{split}
\end{equation}

\begin{equation}
\begin{split}
\mathcal{C}^{\tilde{m}}_{2}(k,k^{\prime}) =  \langle\delta m(k)\delta m(k^{\prime}) \rangle= \Bigg(\frac{A^2 E^2 +B^2 (CE+2D^2 k^2)}{4D (C E + D^2 k^2)}-\frac{A^2E+B^2C}{4DC}l\delta(k)\Bigg)\frac{\delta(k+k^{\prime})}{2\pi},
\end{split}
\label{eq:resft2}
\end{equation}

\begin{equation}
\begin{split}
\mathcal{C}^{\tilde{\rho},\tilde{m}}_{2}(k,k^{\prime}) =  \langle\delta \rho(k)\delta m(k^{\prime}) \rangle=  \frac{ik(B^2C-A^2E)}{4D (C E + D^2 k^2)}\frac{\delta(k+k^{\prime})}{2\pi}.
\end{split}
\label{eq:resft3}
\end{equation}
In real space, Eqs.~\eqref{eq:resft1},~\eqref{eq:resft2} and~\eqref{eq:resft3} transform into
\begin{equation}
\begin{split}
\mathcal{C}^{\rho}_{2}(x,x^{\prime}) = \frac{ \rho (1-\rho)}{L} \big(\delta(x-x^{\prime})-1\big) +  \frac{  \alpha   \rho^2 (1-\rho)^2  (2 \alpha  -\Delta )}{2 D L \sqrt{2 \alpha \Delta \rho (1-\rho)}} e^{-\frac{|x-x^{\prime}|}{\xi}},
\label{eq:res1}
\end{split}
\end{equation}

\begin{equation}
\begin{split}
\mathcal{C}^{m}_{2}(x,x^{\prime}) =  \frac{1}{4L} \big(\delta(x-x^{\prime})-1\big)  -  \frac{  (2\alpha-\Delta) \Delta  \rho (1-\rho)}{16 D L \sqrt{2 \alpha \Delta \rho (1-\rho)}} e^{-\frac{|x-x^{\prime}|}{\xi}},
\end{split}
\label{eq:res2}
\end{equation}

\begin{equation}
\begin{split}
\mathcal{C}^{\rho,m}_{2}(x,x^{\prime}) =  -\text{Sgn}[x-x^{\prime}]\frac{  (2\alpha-\Delta) \rho (1-\rho)}{8 D L }   e^{-\frac{|x-x^{\prime}|}{\xi}},
\end{split}
\label{eq:res3}
\end{equation}
where the correlation length $\xi$ is given by
\begin{equation}
\xi = \frac{D}{\sqrt{2a\Delta\rho_{0}(1-\rho_{0})}}.
\label{eq:corlngth}
\end{equation}
In deriving Eqs.~\eqref{eq:res1} and~\eqref{eq:res2}, we determined the value of $l$ for each equation by requiring that the spatially integrated correlation function be zero, ensuring the conservation of particles and tilts.
\begin{figure}[t!]
\centering
\includegraphics[width=1.0\linewidth] {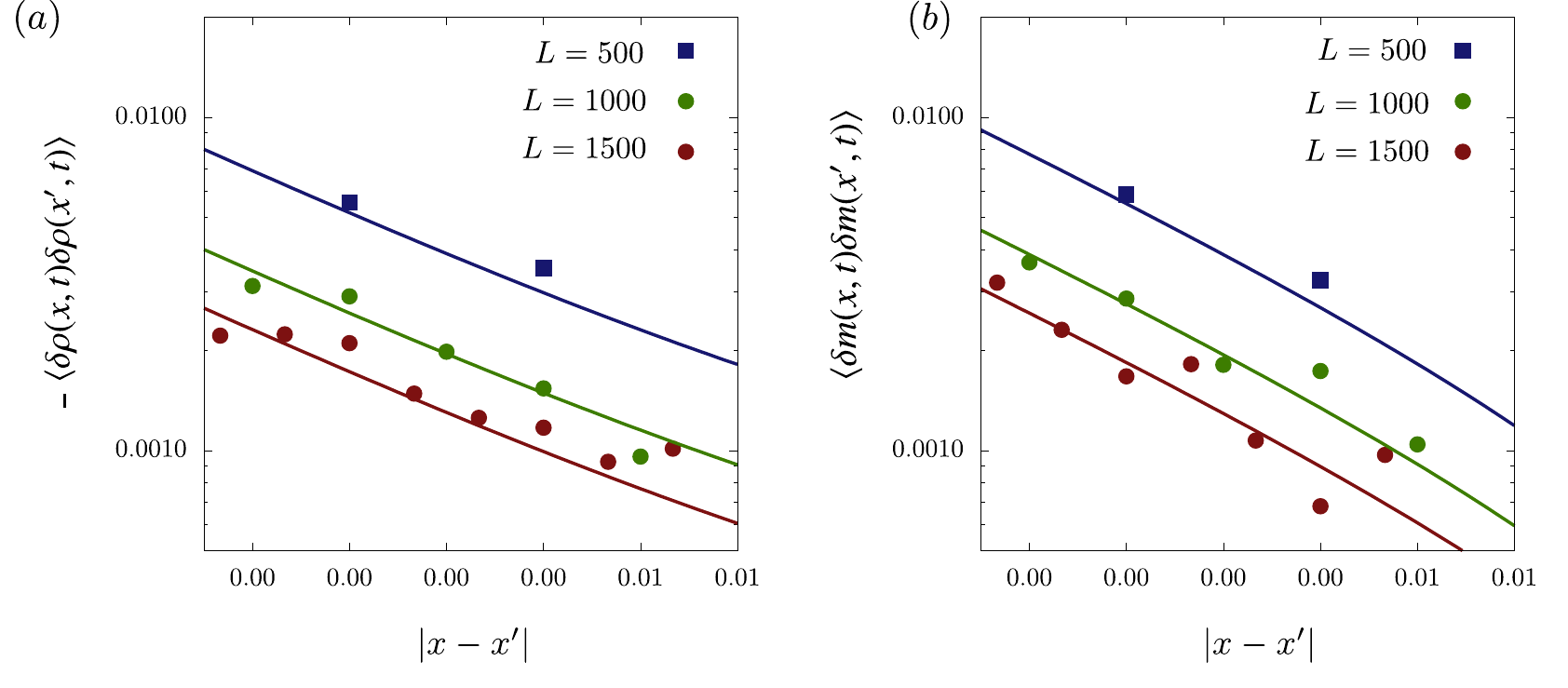}
\caption{Comparison of static correlation functions for fluctuations in ($a$) particle-particle and ($b$) tilt-tilt density fields across multiple system sizes. The parameter values selected are $(\alpha,\beta,\beta^{\prime}) = (140, -180, -180)$. The points represent the predictions generated from simulations, while the lines depict theoretical predictions derived from Eqs.~\eqref{eq:res1} and \eqref{eq:res2}. The systems analyzed have sizes of \(L = 500, 1000, 1500\) and were averaged over 2000 histories.
}
\label{res2}
\end{figure}
The form of the correlation length given in Eq.~\eqref{eq:corlngth} indicates exponentially decaying correlations in the stable region ($\Delta>0$) and oscillating behavior in the linearly unstable regime ($\Delta<0$), where the density cannot be expanded in terms of small fluctuations about a homogeneous steady-state value. Additionally, a detailed analysis of the active lattice gas system also reveals that the theory does not accurately capture the simulations when the correlation length approaches the system size~\cite{jose2023current}. In other words, the correlation function we find for a system holds true away from the critical line, as shown in Figures~\ref{c1},~\ref{res1},~\ref{res2} and~\ref{res3}. The phase diagram of the LH model is constructed using the bias parameters $a, b$ and $b^{\prime}$. Since it is a three-dimensional phase diagram, it will include critical surfaces. To verify the validity of our theory, we first need to identify the locations of these critical planes.
A linear stability analysis of Eq.~\eqref{eq:noiseless} gives, 
\begin{equation}
    [\beta+\beta^{\prime}]_{c} = 0,
\end{equation}
for an infinite system and
\begin{equation}
    [\alpha(\beta+\beta^{\prime})]_{c} = \frac{2 \pi^2 D^{2}}{ L^2 \rho(1-\rho)},
\end{equation}
for a finite system (details in~\ref{sec:appendix A}). This would indicate that, for every $\alpha>0$ plane, there is a critical surface $\beta+\beta^{\prime} = f(L, D, \rho)$, away from which the hydrodynamic theory should work well.

\subsection{Dynamic correlation functions}
\label{sec:dynamic}
\begin{figure}[hbt!]
\centering
\includegraphics[width=1.0\linewidth] {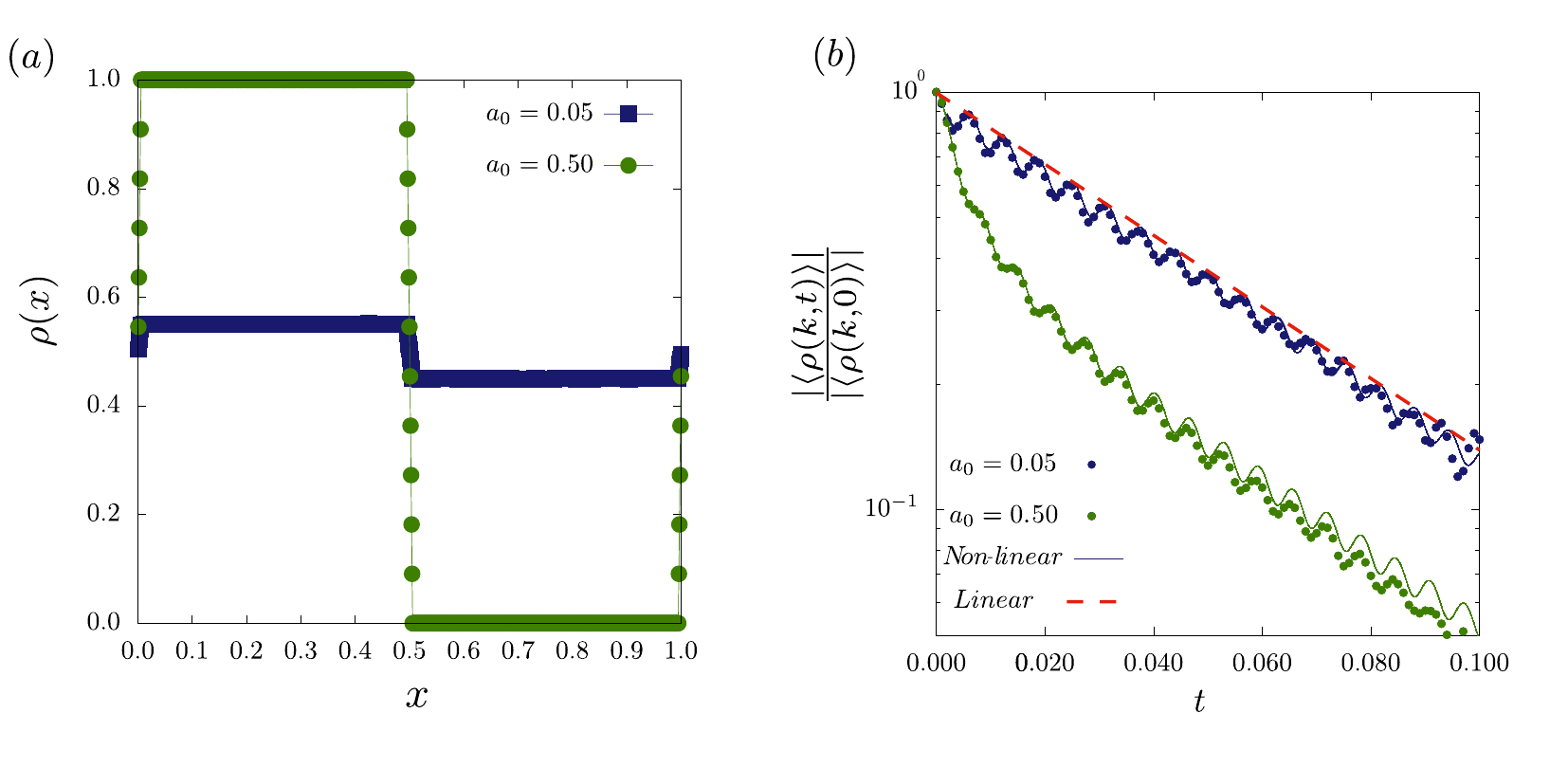}
\caption{(a) Two initial density profiles, allowed to relax to a homogeneous steady-state. The amplitude of the two profiles vary, with $a_{0} =|\rho(x)-\rho_{0}|$ set to $0.05$ and $0.5$. (b) Dynamics of the first Fourier mode of the two density fields as they relax to the steady-state starting from the two initial profiles. Simulation results (points) are plotted against theoretical predictions derived from two approaches: ($i$) the linearized approach (dashed lines) described in Eq.~\eqref{eq:linearrelax}, and ($ii$) numerical solution using fast Fourier transform (solid lines) to the noiseless hydrodynamic equations presented in Eq.~\eqref{eq:noiseless}. Approach ($ii$) can describe the simulation data for both profiles, while approach ($i$) matches the data only when the amplitude is small. A system of size $L=1000$ averaged over $500$ histories is presented.
}
\label{lhrlxtn}
\end{figure}
The linearized theory can be utilized to describe the dynamics of the system, provided that the fluctuations $\delta \rho$ and $\delta m$ in the initial state are not too large. To illustrate this, we examine two initial density profiles: one with a small amplitude and another with a larger amplitude, as shown in Figure~\ref{lhrlxtn}a. We will analyze how these profiles relax to a steady state in the homogeneous region of the phase diagram. According to Eq.~\eqref{eq:lineareqn}, the dynamics of the system under the linear approximation is governed by the eigenvalues of the $\mathcal{M}_{k}$ matrix. These eigenvalues are given by:
\begin{equation}
    \lambda_{\pm} = D k^{2} \pm \sqrt{2 \alpha k^2 \rho_{0} (1-\rho_{0})(\beta+\beta^{\prime})},
\end{equation}
which can be rewritten as,
\begin{equation}
    \lambda_{\pm} = D k^{2} \pm \frac{i D k}{\xi},
\end{equation}
in the homogeneous regime of the parameter space. We can deduce from the complex nature of the eigenvalues in the homogeneous region that kinematic waves, as discussed in references~\cite{Lighthill1, Lighthill2}, are present in this area. The dissipation of the density profile will be determined by the real part of $\lambda_{\pm}$, which is identical for both. Therefore, according to linear theory, we can express this relationship as,
\begin{equation}
\frac{|{\delta \tilde{\rho} (k,t)|}}{|{\delta \tilde{\rho} (k,0)|}} \sim e^{-\lambda t},
\label{eq:linearrelax}
\end{equation}
where $\delta \tilde{\rho}(k,t)$ and $\delta \tilde{\rho}(k,0)$ are complex in nature and $\lambda$ is the real part of either one of the two eigenvalues. In Figure~\ref{lhrlxtn}b, we illustrate the relaxation of the two density profiles.~While the numerical solution to Eq.~\eqref{eq:noiseless} accurately describes the relaxation dynamics in both cases, the linear theory applies only when the fluctuations in densities $\delta \rho$ and $\delta m$ are not too large. This is particularly evident in the case of the initial profile with the larger amplitude, which initially deviates from the linear theory. However, it eventually aligns with the theory once the profile has relaxed sufficiently to reduce the amplitude of the fluctuations.

The success of the linear theory in describing the relaxation dynamics of the system in the homogeneous phase suggests that we can also evaluate dynamic correlations of density fluctuations in this regime. The derivation follows similarly to Section~\ref{sec:derivation-two-point}, replacing the static Fourier transform with
\begin{eqnarray}
\delta\hat{\rho}(k,\omega) = \frac{1}{(2\pi)^2} \int dx dt e^{-i (kx-\omega t)}\delta \rho(x,t),\nonumber\\
\delta\hat{m}(k,\omega) = \frac{1}{(2\pi)^2} \int dx dt e^{-i (kx-\omega t)}\delta m(x,t).
\end{eqnarray}
Thus, Eq.~\eqref{eq:lin1} and Eq.~\eqref{eq:lin2} gives,
\begin{equation}
 \begin{pmatrix} \delta \Tilde{\rho}(k,\omega) \\ \delta \Tilde{m}(k,\omega) \end{pmatrix} = -\mathcal{M}^{-1}_{k,\omega}  \mathcal{R}(k,\omega).
\end{equation}
where
\begin{equation}
   \mathcal{M}_{k,\omega} = \begin{pmatrix} -i\omega +Dk^{2} & iCk\\ iEk & -i\omega +Dk^{2} \end{pmatrix}; \hspace{1.5cm}  \mathcal{R}(k,t) = \begin{pmatrix} Aik\Tilde{\eta_{1}}(k\omega)\\ Bik\Tilde{\eta_{2}}(k,\omega) \end{pmatrix},
   \label{eq:matrix2}
\end{equation}
with $\Tilde{\eta}_{1,2}$ correlated as,
\begin{equation}
\langle \Tilde{\eta}_{i}(k,\omega)  \Tilde{\eta}_{j}(k^{\prime},\omega^{\prime})\rangle =  \frac{\delta_{ij}}{(2\pi)^2}\delta (k+k^{\prime}) \delta (\omega+\omega^{\prime}).  
\end{equation}
The result in Fourier space is given by,
\begin{equation}
\tilde{\mathcal{C}}^{\rho}_{2} = \frac{1}{ L M(k,\omega)} \Biggl( \frac{2 D^{5 }k^{4}}{\Delta^{2}}\frac{1}{\xi^{4}}+ \frac{D^{3}k^{2}}{\alpha \Delta}\frac{1}{\xi^{2}} \Big(D^{2} k^{4}+\omega^{2}\Big)\Biggr) \delta(k+k^{\prime}) \delta(\omega+\omega^{\prime}),
\end{equation}
\begin{equation}
\tilde{\mathcal{C}}^{ m}_{2} = \frac{1}{2 L M(k,\omega)} \Biggl( \frac{ D^{3 }k^{4}\Delta}{2 \alpha}\frac{1}{\xi^{2}}+ D k^{2} \Big(D^{2} k^{4}+\omega^{2}\Big)\Biggr)\delta(k+k^{\prime}) \delta(\omega+\omega^{\prime}),
\label{eq: scaled1}
\end{equation}
\begin{equation}
\tilde{\mathcal{C}}^{\rho,m}_{2} = \frac{1}{ L M(k,\omega)} \Biggl( \frac{{\it i} D^{4} k^{5}}{2 \alpha \Delta \xi^{2}} (2 \alpha-\Delta) + \frac{\omega D^{3} k^{3}}{2 \alpha \Delta \xi^{2}} (2 \alpha+\Delta) \Biggr)\delta(k+k^{\prime}) \delta(\omega+\omega^{\prime}),
\label{eq: scaled2}
\end{equation}
where, 
\begin{equation}
M(k,\omega) = (2\pi)^2\Biggl(\omega^{4} + 2 D^{2}\omega^{2}k^{2} \Big(k^{2}-\frac{1}{\xi^{2}}\Big)+D^{4} k^{4} \Big(k^{2}+\frac{1}{\xi^{2}}\Big)^{2}\Biggr).
\label{eq: scaled3}
\end{equation}

We can use these dynamic correlations to analyze the scaling behavior of the system. This is done using a scaling parameter $b$ to rescale $x \rightarrow b x$ and $t \rightarrow b^{z} t$. Equivalently, we have $k \rightarrow b^{-1} k$ and $\omega \rightarrow b^{-z} \omega$ in Eqs.~\eqref{eq: scaled1}, \eqref{eq: scaled2} or \eqref{eq: scaled3}.
Rescaling $k$ and $\omega$ with $b$ and $b^{z}$ allows us to compute the dynamical scaling exponent $z$. As $b$ increases to larger values, the correlation tends to the asymptotic form,
\begin{equation}
\tilde{\mathcal{C}}^{\rho}_{2} \cong \frac{2 D \rho_{0} (1-\rho_{0})}{L} \Bigg( \frac{k^{2}\omega^{2}+b^{2z-2}D^{2} k^{4}(b^{-2}k^{2}+2\frac{\alpha}{\Delta}\xi^{-2})}{(b^{-z+2}\omega^{2}+b^{z-2}D^{2} k^{2}(k^{2}+b^{2}\xi^{-2}))^{2}+4D^{2}k^{4}\omega^{2}}\Bigg) b^{z+3}\delta(k+k^{\prime}) \delta(\omega+\omega^{\prime}).
\label{eq:scaledasym}
\end{equation}
 As observed in Eq.~\eqref{eq:scaledasym}, the dynamical exponent $z$ extracted from dynamic correlations varies with the chosen length scale.
At length scales smaller than $\xi$, i.e., at criticality, Eq.~\eqref{eq:scaledasym} gives $z=2$. This behavior can be justified numerically by the following argument: Figure~\ref{lhrlxtn}b illustrates the relaxation of the first Fourier mode, described by $e^{-\lambda t}$
where $\lambda = d \Big(k^{2}\pm \frac{{ \it i}k}{\xi}\Big)$. As the system approaches the critical line, the correlation length diverges, and the eigenvalue scales as $\lambda \approx k^{2}$, indicating that the dynamics show diffusive behavior on length scales smaller than the correlation length, as predicted by our scaling analysis.\\

However, we cannot draw a self-consistent dynamical exponent for length scales larger than $\xi$ from \eqref{eq:scaledasym}. This shortcoming may be due to the presence of kinematic waves in the system, which dissipate as criticality is approached. We can observe these waves in the system by looking at the eigenmode fields of the density fluctuations, defined as,
 $\phi_{\pm} = \sqrt{\Delta/2} \delta \rho \pm \sqrt{4 \alpha \rho (1-\rho)} \delta m$. The time evolution of the eigenfields is given by,
 \begin{equation}
 \frac{\partial \phi_{+}}{\partial t}    = D \frac{\partial^{2} \phi_{+}}{\partial x^{2}}  +c_{+}\frac{\partial\delta \phi_{+}}{\partial x}+\frac{\partial \eta_{1}}{\partial x},
 \label{eq:ef1}
 \nonumber
\end{equation}
\begin{equation}
 \frac{\partial \phi_{-}}{\partial t}    = D \frac{\partial^{2}  \phi_{-}}{\partial x^{2}} + c_{-} \frac{\partial  \phi_{-}}{\partial x}+ \frac{\partial  \eta_{2}}{\partial x}.
  \label{eq:ef2}
\end{equation}
Here, $c_{\pm}=\mp \sqrt{2\alpha \Delta \rho (1-\rho)}$ gives the speed of the individual waves. By performing a Galilean shift $x^{'} = x + c_{+}t$, $t^{'}=t$, we can move to a frame that comoves with the $\phi_{+}$ mode. Linearizing and solving for the dynamical correlations of this mode will then yield the value of the dynamical exponent or, equivalently, describe how the wave dissipates. We outline some of the key points of the calculation in the following. The time evolution equation in the frame of $\phi_{+}$ is described by,
 \begin{equation}
 \frac{\partial \phi_{+}}{\partial t}    = D \frac{\partial^{2} \phi_{+}}{\partial x^{2}}  +\frac{\partial \eta_{1}}{\partial x},
 \label{eq:ef3}
 \nonumber
\end{equation}
\begin{equation}
 \frac{\partial \phi_{-}}{\partial t}    = D \frac{\partial^{2}  \phi_{-}}{\partial x^{2}} + (c_{-} - c_{+}) \frac{\partial  \phi_{-}}{\partial x}+ \frac{\partial  \eta_{2}}{\partial x}.
  \label{eq:ef4}
\end{equation}
By solving for the modes, one can determine the dynamic correlation of the $\phi_{+}$ field to be given by,
\begin{equation}
   \tilde{\mathcal{C}}^{\phi_{+}}_{2} \cong  \frac{k^{2}}{D^{2} k^{2}+ \omega^{2}},
  \label{eq:ef5}
\end{equation}
which on rescaling becomes,
\begin{equation}
   \tilde{\mathcal{C}}^{\phi_{+}}_{2} \cong  \frac{k^{2}}{b^{2z-4}D^{2} k^{2}+ \omega^{2}}.
  \label{eq:ef6}
\end{equation}
If we move to a frame that comoves with the $\phi_{-}$ mode, we will see that the dynamic correlation of the $\phi_{-}$ mode also gives $z=2$.  Thus, we find that the linear theory predicts that the wave dissipation follows the Edwards-Wilkinson exponent~\cite{Barabasi_Stanley_1995}. This aligns with previous observations made in the same regime using a symmetry argument~\cite{dasbasu_2001}. However, we recognize that our linearized theory has limitations in this regard. The linear analysis does not take into account the impact of higher-order fluctuations in the density fields on $z$. For example, previous studies~\cite{dasbasu_2001} revealed possible logarithmic corrections in the behavior of the correlation functions caused by these higher-order fluctuations in density fields.
\section{Phase transition within the linearly unstable region}
\label{sec: phase transition}
\begin{figure}[t!]
\centering
\includegraphics[width=1.0\linewidth] {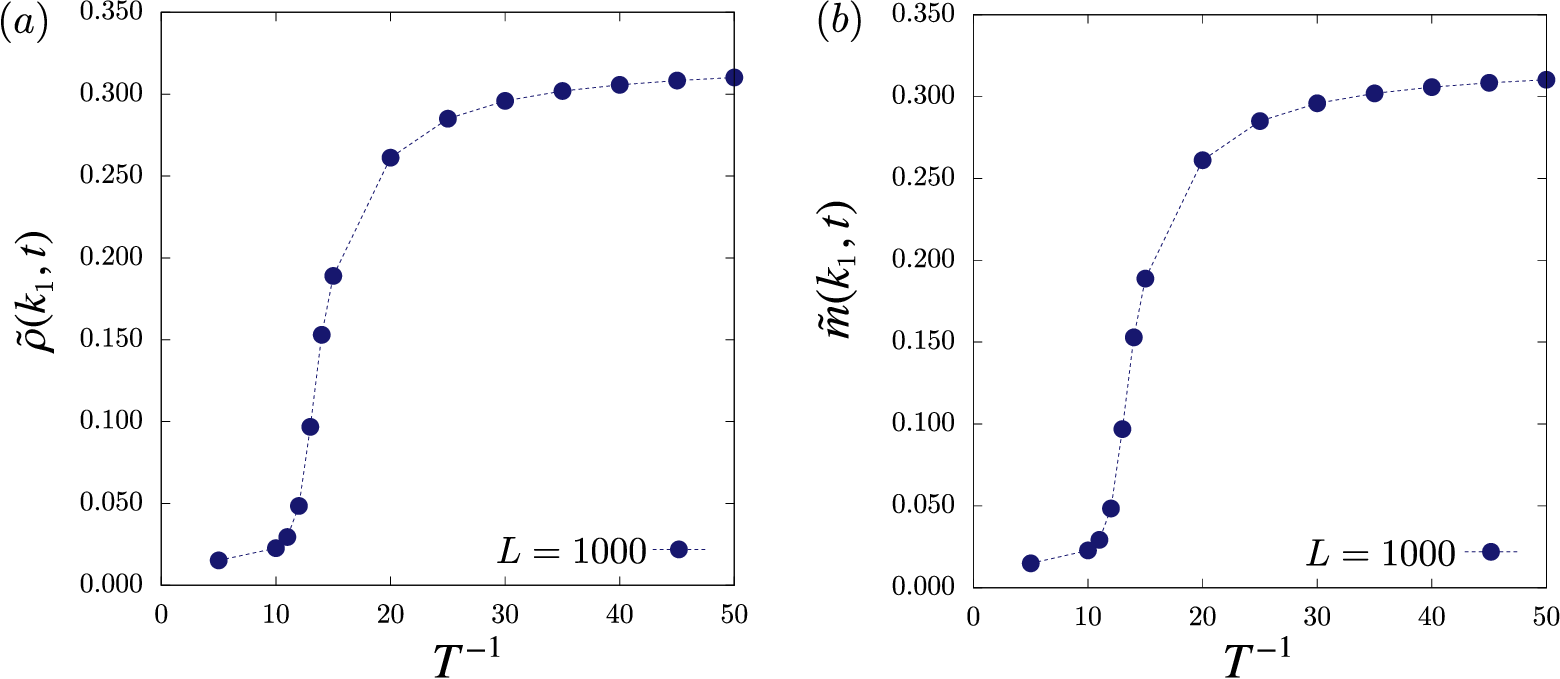}
\caption{Behavior of the first Fourier mode of ($a$) particle density ($\tilde{\rho}(k_{0},t)$) and ($b$) tilt density ($\tilde{m}(k_{0},t)$) along a locus in the phase diagram that obeys detailed balance ($a = b = b^{\prime}$) with respect to the Hamiltonian described in Eq. \eqref{eq:hamil} ($\lambda = \rho_{0} = 0.5$). The numerical data (line points) indicate a continuous phase transition from a disordered phase to an ordered phase as the control parameter, $T^{-1}$, is increased. In Ref.~\cite{Chakraborty_2017}, Chakraborty {\it et al.}~identified a system size independent $T_{c}$ (Eq.~\eqref{eq:crittemp}) for this transition. This $T_{c}$ corresponds to a system size dependent point in the phase diagram ($a_c,b_c,b_c^{\prime}$) where $\sqrt{a_c^{2}+b_c^{2}+{b_{c}^{\prime}}^{2}} \sim 1/L$. Consequently, the critical point is shifted by $\mathcal{O}(1/L)$ from the origin in the sLH model. A system of size $L = 1000$ is used for the simulation.} 
\label{chak1}
\end{figure}
Linear stability analysis points to an unstable region in the subspace $\beta+\beta^{\prime}>0;~\alpha>0$ of the parameter space of the LH model, as shown in Figure~\ref{pic2}c. This is supported by the numerical observation that a phase separation appears when $\alpha, \beta, \beta^{\prime} > 0$. However, Chakraborty~{\it et al.} have already established that in the hydrodynamic limit, a phase transition is observed between an ordered and a disordered phase in the scaled LH model within this linearly unstable regime~\cite{Chakraborty_2017}. Although the observation is general to the subspace $\beta+\beta^{\prime}>0;~\alpha>0$, their theoretical argument is limited to loci within this subspace, where the rates obey detailed balance with respect to a given Hamiltonian. The Hamiltonian is given by
\begin{equation}
    \mathcal{H} = \sum_{i=1}^{L} (n_{i}-\lambda)h_{i},
    \label{eq:hamil}
\end{equation} 
where, $\lambda = \rho$, the total density of the heavy particles. With respect to this Hamiltonian, the detailed balance condition is satisfied for the following choice of rates,
\begin{equation}
    \frac{D-a}{D+a} = e^{-1/TN}, \hspace{1cm} \frac{E-b}{E+b} = e^{-(2-2\lambda)/TN}, \hspace{1cm} \frac{E-b^{\prime}}{E+b^{\prime}} = e^{-2\lambda/TN}.
\end{equation}
Such a definition makes the energy extensive and thereby comparable to the entropy. This leads to a phase transition at 
\begin{equation}
    \frac{1}{T_{c}} = \frac{2 \pi}{\sqrt{\lambda(1-\lambda)}}.
    \label{eq:crittemp}
\end{equation}
Thus, it was found that for every point in the parameter space given by a particular value of $\lambda$, there is a critical point that separates the ordered and disordered regime. This point is demonstrated for the case where $\lambda = 0.5$ in Figure~\ref{chak1}. Although this formalism gives us critical values along a certain locus that obeys detailed balance, the existence of the transition is general to the subspace $\beta+\beta^{\prime}>0;~\alpha>0$, with the entire region divided into disordered and ordered states. 

The phase transition in the scaled LH model is therefore in contradiction to the prediction of the linear stability analysis, which predicts that the entire region above the $b+b^{\prime}$ line, is unstable. However, this can be easily established as a finite-size effect. We see that on increasing the system size, the distance to the theoretically predicted critical point from the origin for the different paths that obey detailed balance in the parameter space falls as $x_{c} = \sqrt{a_{c}^{2}+b_{c}^{2}+b_{c}^{\prime2}} \sim 1/L$, coinciding with the prediction of the linear stability analysis in the infinite size limit.

The ordered state that emerges after the phase transition in the scaled model differs from the unscaled model, as mentioned in Section~\ref{sec:model}.~This new state is no longer characterized by a strong phase separation of the particles. Instead, we observe a normal phase separation of both species, which leads to the loss of phases such as strong phase separation (SPS), infinitesimal current phase (IPS), and finite current phase separation (FPS), among others. 

To understand why this happens, we return to an analysis of the unscaled model. When the rates $\alpha,\beta,\beta'$ are of $\mathcal{O}(1)$, there are three different ordered phases, all of which show strong phase separation of the particles and varying degrees of order in the tilt arrangement. By strong phase separation, one essentially refers to the absence of droplets of opposite species far from the interface in a cluster. In an earlier study, Lahiri {\it et al.}~\cite{Lahiri_2000} explored the broadening of the interface found in the system along a locus where detailed balance holds ($a=b=b^{'}$), and the steady state measure of the system is given by $\sim e^{- \mathcal{H}/T}$ as the temperature, $T$ is raised, with the Hamiltonian given by
\begin{equation}
\mathcal{H} = \epsilon \sum_{k=1}^{N}h_{k}\sigma_{k}.
\end{equation}
Here, the temperature can be related to the rates as
\begin{equation}
\frac{\epsilon}{T}= \frac{1}{2} \ln \left( \frac{D+a}{D-a} \right),
\end{equation}
and density of either species of particle, $\rho = 0.5$. The analysis predicted that at a temperature $T$, 
\begin{equation}
\langle\sigma_{k}\rangle = \tanh \Bigl(\frac{\epsilon}{T} (k-k_{0})\Bigr),
\end{equation}
where $k_{0}$ is the site at which the interface occurred at $T=0$. The conclusion was that the width of the interface, $w$, at a temperature $T$, namely the region where $\langle\sigma_{k}\rangle$ deviates substantially from $1$, is given by,
\begin{equation}
w \sim \frac{T}{\epsilon}.
\end{equation}
At finite temperatures, the interface broadens to a finite number of sites, leaving exponentially low chances for islands of opposite species away from the interface. The same arguments help motivate the disappearance of the different ordered phases in transitioning to the scaled model. In the scaled model, we are considering a temperature of the order of the system size. Consequently, the interface width would be of the order of system size, ruling out strong phase separation.

\begin{figure}[t!]
\centering
\includegraphics[width=1.0\linewidth] {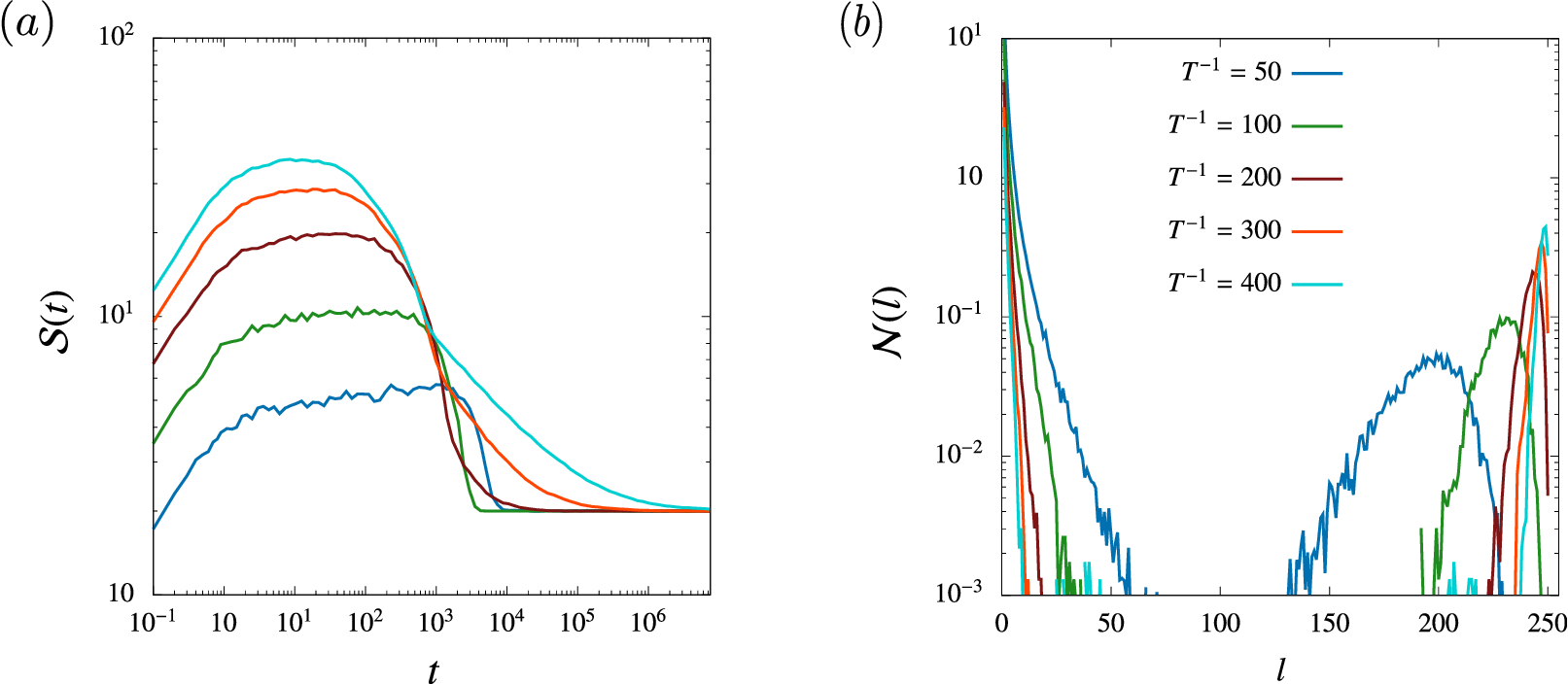}
\caption{(a) Comparison of the behavior of $\mathcal{S}$ for different values of the parameters as the system evolves to steady-state starting from a Gaussian initial state. The parameters ($a, b, b^{\prime}$) are selected along a path in the phase diagram that obeys detailed balance ($a = b = b^{\prime}$) with respect to the Hamiltonian described in Eq.~\eqref{eq:hamil} ($\lambda = \rho_{0} = 0.5$) using inverse temperature $T^{-1}$ as the control parameter. The specific values chosen for $T^{-1}$ are 50, 100, 200, 300, and 400. (b) The corresponding steady-state cluster size distribution. The broad distribution observed when $T^{-1} \ll \mathcal{O}(L)$ indicates that there is no strong phase separation in this regime. As the control parameter increases to $\mathcal{O}(L)$, the distribution peak sharpens around $l \sim L/2$. The system size used here is $L = 500$. }
\label{cluster}
\end{figure}

Parameters such as cluster size distribution and correlation function help us identify this change in the phase diagram. Figure~\ref{cluster} shows the change in the distribution of the cluster size, as well as the configuration-averaged cross-correlation in a system of size $L = 500$ with the control parameter $T^{-1}$ that varies from $\mathcal{O}(1)$ to $\mathcal{O}(L)$. 
The cross-correlation function $S(t)$, defined in \cite{Mahapatra2020Light}, is a local variable that evaluates the correlation between one species and the gradient of the other. For example,
\begin{equation}
S_{\sigma \Delta \tau} = \sum_{j=1}^{L} \frac{1}{2} (\tau_{j-\frac{1}{2}} - \tau_{j+\frac{1}{2}}) \sigma_{j},
\end{equation}
counts triads of the kind $/\circ \setminus$ and $\setminus\bullet/$ as 1 and $/\bullet \setminus$ and $\setminus\circ/$ as -1. Here, $L$ represents the size of the system. An alternative definition for $S(t)$, which gives the same result as this under periodic conditions, is provided by,
\begin{equation}
S_{\tau \Delta \sigma} = \sum_{j=1}^{L} \frac{1}{2} (\sigma_{j+1} - \sigma_{j})\tau_{j + \frac{1}{2}}.
\end{equation}
The configuration-averaged version $\mathcal{S}$, which we sample, gives the average correlation of the same across the lattice. It is defined as, 
\begin{equation}
\mathcal{S}(t) = \frac{1}{L}\langle S_{\sigma \Delta \tau}\rangle = \frac{1}{L}\langle S_{\tau \Delta \sigma}\rangle.
\end{equation}
Referring back to Figure~\ref{cluster}, we see that in the steady state (indicated by the constant region in the configuration-averaged cross-correlation function), there is a shift in the distribution of the cluster towards the right end of cluster sizes on increasing $T^{-1}$. As $T^{-1}$ becomes of the order of system size, $\mathcal{S}$ is expected to fall logarithmically in time, and the cluster size distribution would be concentrated at  $l\sim L/2$, indicating SPS.

Finally, we also note that the scaled model shows a continuous phase transition, as shown in Figure~\ref{chak1} and in a previous study of the same model~\cite{Chakraborty_2017}. In contrast, our preliminary numerical studies indicate a mixed-order transition~\cite{Mukamel_2024} for the unscaled model. The order parameter displays a discontinuous jump, while the correlation length appears to diverge when approaching the FDPO line from the disordered side.

\section{Discussion}
\label{sec:discussion}    
In this paper, we have derived the exact fluctuating hydrodynamics for the scaled Light-Heavy (sLH) model, where the microscopic rates are scaled by the system size. This scaled model differs from the unscaled model in several significant ways. While the unscaled model exhibits multiple phases and transitions, the scaled model reveals only a single transition to one ordered phase. Additionally, the strong phase separation seen in the unscaled model is absent in the scaled model, resulting in only normal phase separation. Moreover, the Fluctuation-Dominated Phase Ordering (FDPO) along the separatrix is also absent in the scaled model.

We used this fluctuating hydrodynamics framework to obtain the steady-state correlations in the homogeneous phase. Our theory predicts a diverging correlation length as the line separating the ordered and disordered phases is approached. These theoretical results show an excellent match with microscopic simulations. While the fluctuating hydrodynamics framework has been used successfully to study models such as the active lattice gas~\cite{agranov2021exact} and the ABC model~\cite{Clincy_2003, Bodineau_2011}, the LH model differs from these, in the form of the noise terms as well as the correlation function. It would be interesting to extend our study to the critical line itself, and in particular, to check whether the correlation function displays a power-law behavior at the critical point. It would also be interesting to test whether the LH model with unscaled rates (uLH) also displays a diverging correlation length. Our preliminary numerical simulations suggest that indeed such a diverging correlation length does seem to appear in the unscaled model as well; however, an analytical description for this behavior is presently lacking. Given that the order parameter jumps from $0$ to $1$ across the critical line, this diverging correlation length in the unscaled model could be indicative of the mixed nature of the transition between ordered and disordered phases separated by the Fluctuation-Dominated Phase Ordering (FDPO) phase that is known to occur in this model~\cite{Lahiri_2000,das2001fluctuation}, and therefore would be very interesting to study further.

We have also predicted the dynamical exponents of the scaled system using a linearized theory based on fluctuating hydrodynamics. The unscaled model is known to display diverse dynamical behaviors, with exponents such as diffusive, KPZ, modified KPZ, as well as $\frac{5}{3}$-Levy~\cite{Chakraborty_2019} appearing in different regimes. While our investigation of the homogeneous phase of the scaled model has revealed only diffusive behavior so far, exploring the different possible dynamical regimes in the scaled model presents a promising direction to explore further.
\section{Acknowledgments}
We thank Rashmiranjan Bhutia and Stephy Jose for useful discussions. We also acknowledge useful discussions with Anupam Kundu related to the action formalism. M.B. acknowledges the support of the Indian National Science Academy (INSA). We acknowledge the support of the Department of Atomic Energy, Government of India, under Project Identification No. RTI4007.

\appendix
\section{}
\section*{Linear stability analysis for the LH Model}
\label{sec:appendix A}
We have the hydrodynamic expressions given by,
\begin{equation}
 \frac{\partial \rho}{\partial t}    = D \frac{\partial^{2} \rho}{\partial x^{2}} +\frac{\partial}{\partial x} \bigg[2\alpha\rho(1-\rho)(1-2m)\bigg], 
\end{equation}
\begin{equation}
\frac{\partial m}{\partial t}    = D \frac{\partial^{2} m}{\partial x^{2}} +\frac{\partial}{\partial x} \bigg[m(1-m)(2\rho(\beta+\beta^{\prime})-2\beta^{\prime})\bigg] .
\label{eq:yetryejh}
\end{equation}
In the hydrodynamic limit, we can obtain the solutions for the two-point correlation function by linearising these equations for small fluctuations $\rho(x,t) = \rho_{0} + \delta  \rho(x,t)/\sqrt{L}$ and $m(x,t) = m_{0} + \delta m(x,t)/\sqrt{L}$. This gives,
 \begin{equation}
 \frac{\partial \delta \rho}{\partial t}    = D \frac{\partial^{2} \delta \rho}{\partial x^{2}}   -4\alpha\rho_{0}(1-\rho_{0})\frac{\partial\delta m}{\partial x} + 2\alpha(1-2\rho_{0})(1-2m_{0})\frac{\partial\delta \rho}{\partial x},
\end{equation}
\begin{equation}
 \frac{\partial \delta m}{\partial t}    = D \frac{\partial^{2} \delta m}{\partial x^{2}} +2\Delta m_{0}(1-m_{0})\frac{\partial\delta \rho}{\partial x} + (1-2m_{0})(-2\rho_{0}\Delta + \beta^{\prime})\frac{\partial\delta m}{\partial x} .
\end{equation}
We can further simplify the expression by taking into account the fact that we are looking at systems with no net tilt on the surface, i.e., $m_{0} = \frac{1}{2}$. Then
 \begin{equation}
 \frac{\partial \delta \rho}{\partial t}    = D \frac{\partial^{2} \delta \rho}{\partial x^{2}}   -4\alpha\rho_{0}(1-\rho_{0})\frac{\partial\delta m}{\partial x},
\end{equation}
\begin{equation}
 \frac{\partial \delta m}{\partial t}    = D \frac{\partial^{2} \delta m}{\partial x^{2}} -\frac{\Delta}{2} \frac{\partial \delta \rho}{\partial x}.
\end{equation}
Here, the linear term in the expression makes it difficult to solve it in real space, and therefore we analyze it in Fourier space. Using continuum Fourier transforms given by,
\begin{equation}
	\delta\tilde{m}\left(k,t \right)= \frac{1}{2 \pi} \int dx~ e^{-i kx }\delta m\left(x,t\right)
	\quad;\quad
	\delta\tilde{\rho}\left(k,t \right)= \frac{1}{2 \pi} \int dx~ e^{-ikx}\delta\rho\left(x,t\right),
 \label{eq:FourierTransform}
\end{equation}
we derive the system of linear equations
\begin{equation}
 \partial_{t} \begin{pmatrix} \delta \Tilde{\rho}(k,t) \\ \delta \Tilde{m}(k,t) \end{pmatrix} = \mathcal{M}_{k} \begin{pmatrix} \delta \Tilde{\rho}(k,t) \\ \delta \Tilde{m}(k,t) \end{pmatrix}.
\end{equation}
Here, the matrix $\mathcal{M}_{k}$ is 
\begin{equation}
   \mathcal{M}_{k} = \begin{pmatrix} -Dk^{2} & -iCk\\ -iEk & -Dk^{2} \end{pmatrix};
\end{equation}
with $C=4\alpha\rho_{0}(1-\rho_{0})$ and $E = -\frac{\beta+\beta^{\prime}}{2}$, and where $\Tilde{\eta}_{1,2}$ are Gaussian white noises.
The eigenvalues are given by, 
\begin{equation}
    \lambda = -D k^{2} \pm \sqrt{2 \alpha k^2 \rho_{0} (1-\rho_{0})(\beta+\beta^{\prime})}
\end{equation}
which gives,
\begin{equation}
\begin{pmatrix} \delta \Tilde{\rho}(k,t) \\ \delta \Tilde{m}(k,t) \end{pmatrix}
= e^{\lambda_{1}t}\ket{\lambda_{1}}\bra{\lambda_{1}}\begin{pmatrix} \delta \Tilde{\rho}(k,0) \\ \delta \Tilde{m}(k,0) \end{pmatrix} + e^{\lambda_{2}t}\ket{\lambda_{2}}\bra{\lambda_{2}}\begin{pmatrix} \delta \Tilde{\rho}(k,0) \\ \delta \Tilde{m}(k,0) \end{pmatrix}.
\end{equation}
Here, we can obtain the unstable regime by checking for the region where the solution blows up. This condition is satisfied by,
\begin{equation}
    \lambda = -D k^{2} + \sqrt{2 \alpha k^2 \rho_{0} (1-\rho_{0})(\beta+\beta^{\prime})} > 0
\end{equation}
 We further filter this result by looking for the region where for the lowest mode, $k = 0$ is unstable. This is a sufficient condition for stability because if the solution is unstable for the lowest mode, then naturally, it will be unstable for the higher modes. Thus,
\begin{equation}
    [(\beta+\beta^{\prime})] > 0
\end{equation}
The regions where the lowest mode is stable but not the highest one, that is, $k \ne 0$,
\begin{equation}
    [\alpha  (\beta+\beta^{\prime})] > \frac{2 \pi^{2}D^2}{L^2 \rho_{0} (1-\rho_{0})}
\end{equation}
gives the region of instability for a finite system.

\section{}
\section*{Field-theoretic formulation of the LH model}
\label{sec:appendix b}
In this section, we present the path integral formulation used to derive the fluctuating hydrodynamics of the LH model. Details of the model and its dynamics can be found in Section~\ref{sec:model}. We consider a system composed of two sublattices, each containing  $L$ sites, indexed by $i$ and $i + \frac{1}{2}$, respectively. In this setup, two sites within the same sublattice are separated by a distance $\epsilon$, i.e., $x_{i+1} - x_{i} = \epsilon$.  The overall size of the system is defined as $\mathcal{L}=L \epsilon$. The system evolves in discrete time intervals of $\delta$, accumulating to a total time $T$ over $N$ steps. We index different time steps with $j$.

Variables $\eta^{\sigma,\tau}_{i}(t_{j})$ ,which can take the values $0, 1$ are introduced to distinguish different configuration. Additionally, we define a variable $J_{i}(t_{j})$ for particles and tilts, which describes their evolution at a time $t_{j}$. This quantity is given by, $J^{\sigma,\tau}_{i}(t_{j}) = \eta^{\sigma,\tau}_{i}(t_{j+1}) - \eta^{\sigma,\tau}_{i}(t_{j}) $.
A trajectory of the system is then determined by the set $\{\eta\}$ containing all $\eta^{\sigma,\tau}_{i}(t_{j}) $. The path integral for probability $P(\{\eta\})$ of observing a given trajectory ${\eta}$ of particles and tilts is then expressed as, 
\begin{equation}
P(\{\eta\}) = \Bigg \langle \prod_{i=1}^{L} \prod_{j=1}^{N} \delta \big(\eta^{\sigma}_{i}(t_{j+1}) - \eta^{\sigma}_{i}(t_{j}) - J^{\sigma}_{i}(t_{j})\big) \delta \big(\eta^{\tau}_{i}(t_{j+1}) - \eta^{\tau}_{i}(t_{j})-J^{\tau}_{i}(t_{j})\big) \Bigg \rangle _{\{J\}},
\end{equation}
which in the integral formulation of the Dirac delta function translates to,
\begin{equation}
\begin{split}
P(\{\eta\}) = \int \prod_{J=1}^{N} \Bigg[ \prod_{i=1}^{L} \Bigl[ d\hat{\rho}_{i}(t_{j}) d\hat{m}_{i}(t_{j})  e^{-\hat{\rho}_{i}(t_{j})\big[\eta^{\sigma}_{i}(t_{j+1}) - \eta^{\sigma}_{i}(t_{j})\big]- \hat{m}_{i}(t_{j}) \big[\eta^{\tau}_{i}(t_{j+1}) - \eta^{\tau}_{i}(t_{j})\big]} \Bigr]\\
 \Bigg \langle \prod_{i=1}^{L} \Bigl[e^{\hat{\rho}_{i}(t_{j}) J^{\sigma}_{i}(t_{j})+\hat{m}_{i}(t_{j}) J^{\tau}_{i}(t_{j})}\Bigr]\Bigg \rangle _{\{J\}} \Bigg].
 \label{eq:probdistricopy}
\end{split}
\end{equation}
Here, $\hat{\rho}$ and $\hat{m}$ represent the conjugate fields and $\langle.\rangle_{\{J^{j}\}}$ is the average over all configurations $\mathcal{C}$ in $\{J^{j}\}$.
Denoting $f(C) = \prod_{i=1}^{L} \Bigl[e^{\hat{\eta}^{\sigma}_{i}(t_{j}) J^{\sigma}_{i}(t_{j})+\hat{\eta}^{\tau}_{i}(t_{j}) J^{\tau}_{i}(t_{j})}\Bigr]$
we get,
\begin{equation}
\Bigg \langle \prod_{i=1}^{L} \Bigl[e^{\hat{\eta}^{\sigma}_{i}(t_{j}) J^{\sigma}_{i}(t_{j})+\hat{\eta}^{\tau}_{i}(t_{j}) J^{\tau}_{i}(t_{j})}\Bigr]\Bigg \rangle _{\{J\}} = \sum _{C\in \{J^{j}\}} f(\mathcal{C}) P(\mathcal{C}|\{\eta^{j}\}),
\label{eq:fc}
\end{equation}
where, $P(\mathcal{C}|\{\eta^{j}\})$ is the probability to observe a configuration $\mathcal{C}$ given by the set $\{\eta^{j}\}$ at the time $t_{j}$. Eq.~\eqref{eq:fc} can be adapted for the LH model by incorporating its local update rules as given in Figure~\ref{pic2}. 
\begin{equation}
    P(\{\eta\})  = \int \prod_{J=1}^{N} \prod_{i=1}^{L} \mathcal{D}[\hat{\rho},
    \hat{m}] e^{\mathcal{S}},  
\end{equation}
where,
\begin{equation}
    \mathcal{S}  = \sum_{i,j}
    \left(
    \begin{aligned}
        &\hat{\rho}(x_{i},t_{j}) \big(\eta^{\sigma}(x_{i},t_{j}+\delta) - \eta^{\sigma}(x_{i},t_{j})\big) + 
        \hat{m}(x_{i},t_{j}) \big(\eta^{\tau}(x_{i},t_{j}+\delta) - \eta^{\tau}(x_{i},t_{j})\big) +\\
        &\big(e^{\hat{\rho}(x_{i},t_{j})-\hat{\rho}(x_{i}+\epsilon,t_{j})}-1\big)\eta^{\sigma}(x_i,t_{j}) \big(1-\eta^{\sigma}(x_i+\epsilon,t_{j})\big)\big(1-\eta^{\tau}(x_{i},t_{j})\big) \big(D+a\big) \delta+\\  
         &\big(e^{-\hat{\rho}(x_{i},t_{j})+\hat{\rho}(x_{i}+\epsilon,t_{j})}-1\big)\eta^{\sigma}(x_i+\epsilon,t_{j}) \big(1-\eta^{\sigma}(x_i,t_{j})\big)\eta^{\tau}(x_{i},t_{j}) \big(D+a\big) \delta+\\           
         &\big(e^{\hat{\rho}(x_{i},t_{j})-\hat{\rho}(x_{i}+\epsilon,t_{j})}-1\big)\eta^{\sigma}(x_i,t_{j}) \big(1-\eta^{\sigma}(x_i+\epsilon,t_{j})\big) \eta^{\tau}(x_{i},t_{j})\big(D-a\big) \delta+\\   
         &\big(e^{-\hat{\rho}(x_{i},t_{j})+\hat{\rho}(x_{i}+\epsilon,t_{j})}-1\big)\eta^{\sigma}(x_i+\epsilon,t_{j}) \big(1-\eta^{\sigma}(x_i,t_{j})\big)\big(1-\eta^{\tau}(x_{i},t_{j})\big) \big(D-a\big) \delta+\\
         &\big(e^{\hat{m}(x_{i}-\epsilon,t_{j})-\hat{m}(x_{i},t_{j})}-1\big)\eta^{\tau}(x_i-\epsilon,t_{j}) \big(1-\eta^{\tau}(x_i,t_{j})\big)\eta^{\sigma}(x_{i},t_{j}) \big(D+b\big) \delta+\\
         &\big(e^{-\hat{m}(x_{i}-\epsilon,t_{j})+\hat{m}(x_{i},t_{j})}-1\big)\big(1-\eta^{\tau}(x_i-\epsilon,t_{j})\big) \eta^{\tau}(x_i,t_{j})\eta^{\sigma}(x_{i},t_{j}) \big(D-b\big) \delta+\\
         &\big(e^{\hat{m}(x_{i}-\epsilon,t_{j})-\hat{m}(x_{i},t_{j})}-1\big)\eta^{\tau}(x_i-\epsilon,t_{j}) \big(1-\eta^{\tau}(x_i,t_{j})\big)\big(1-\eta^{\sigma}(x_{i},t_{j})\big) \big(D-b^{\prime}\big) \delta+\\
         &\big(e^{-\hat{m}(x_{i}-\epsilon,t_{j})+\hat{m}(x_{i},t_{j})}-1\big)\big(1-\eta^{\tau}(x_i-\epsilon,t_{j})\big) \eta^{\tau}(x_i,t_{j})\big(1-\eta^{\sigma}(x_{i},t_{j})\big) \big(D+b^{\prime}\big) \delta
    \end{aligned}
    \right)
\end{equation}
At this stage, we would like to transition to continuum space. To accomplish this, we will coarse-grain the system by using boxes of length $\Delta << \mathcal{L}$ as,
\begin{align}
    \rho(x_{i},t_{j}) &= \frac{1}{\Delta} \sum\limits_{x_i = x_i}^{x_i + \Delta} \eta^{\sigma}(x_{i},t_{j}) \\
    m(x_{i},t_{j}) &= \frac{1}{\Delta} \sum\limits_{x_i = x_{i+\frac{1}{2}}}^{x_{i+\frac{1}{2}} + \Delta} \eta^{\tau}(x_{i},t_{j}) 
\end{align}
This allows us to perform a Taylor expansion up to the order of $\mathcal{O}(\epsilon^{2})$ and $\mathcal{O}(\delta)$. Additionally, we can transition from the summation of time and space variables to integration, as we are interested in the limit  $\epsilon,\delta\rightarrow0$. Thus, we get
\begin{equation}
    \mathcal{S}  = \frac{1}{\epsilon}\int^{\mathcal{L}}_{0}\int^{T}_{0}  S[\rho,m,\hat{\rho},\hat{m}] dx dt
\end{equation}
Rescaling space as $x\rightarrow x/\mathcal{L}$ and time as $t\rightarrow t/L^{2}$, we arrive at the final expression for the action of the system in the diffusive limit,
\begin{equation}
\begin{split}
S = \dot{\rho} \hat{\rho} + \dot{m} \hat{m} +  D \hat{\rho}_{x} \rho_{x} + D \hat{m}_{x} m_{x} +
D\rho (1-\rho)\hat{\rho}_{x}^{2} + Dm (1-m)\hat{m}_{x}^{2} \\+ 2 \alpha \rho (1-\rho) (1-2m)\hat{\rho}_{x} + 
2 m(1-m)\big((\beta+\beta^{\prime})\rho- \beta^{\prime}\big) \hat{m}_{x}.
\end{split}
\end{equation}
Here, $\alpha, \beta, \beta^{\prime}$ are the rescaled rates, i.e., $\alpha =a L$, $\beta =b L$ and $\beta^{\prime} =b^{\prime} L$.
As we are in the continuum limit of space $(\epsilon\rightarrow 0)$, we must perform a saddle-point analysis of this action to derive the evolution equations for $\rho$ and $m$. This gives us the expressions,
\begin{equation}
 \frac{\partial \rho}{\partial t}    = D \frac{\partial^{2} \rho}{\partial x^{2}} +\frac{\partial}{\partial x} \bigg[2\alpha\rho(1-\rho)(1-2m)\bigg] + \frac{\partial [2D \rho (1-\rho)\hat{\rho}_{x}]}{\partial x},
 \nonumber
\end{equation}
\begin{equation}
\frac{\partial m}{\partial t}    = D \frac{\partial^{2} m}{\partial x^{2}} +\frac{\partial}{\partial x} \bigg[2m(1-m)(\rho(\beta+\beta^{\prime})-\beta^{\prime})\bigg] +  \frac{\partial [2D m (1-m)\hat{m}_{x}]}{\partial x}, \label{eq:partdiffaaction2}
\end{equation}
as stated in Eq.~\eqref{eq:partdiffaaction3} mentioned in Section~\ref{sec:Action-LH}.
\section*{References}
\bibliographystyle{iopart-num}
\bibliography{lh}

\end{document}